\documentclass[5p]{elsarticle}

\usepackage{float}
\usepackage{lineno,hyperref}
\usepackage{subcaption}
\usepackage[toc,page]{appendix}
\journal{The future of astronomical data formats}

\usepackage{algorithm}
\usepackage{algorithmic}
\usepackage[labelfont=bf,justification=raggedright,singlelinecheck=false]{caption}
\usepackage{graphicx}

\begin{document}

\begin{frontmatter}

\title{Data compression for the First G-APD Cherenkov Telescope}

\newcommand{\uniw}{$^{a}$}
\newcommand{\ethz}{$^{b}$}
\newcommand{\unige}{$^{c}$}
\newcommand{\rwth}{$^{^e}$}
\newcommand{\tudo}{$^{d}$}

\author[ethz]{M.~L.~Ahnen}
\author[unige]{M.~Balbo}
\author[uniw]{M.~Bergmann}
\author[ethz]{A.~Biland}
\author[ethz,rwth]{T.~Bretz\corref{cor1}}
\author[tudo]{J.~Bu\ss}
\author[uniw]{D.~Dorner}
\author[tudo]{S.~Einecke}
\author[tudo]{J.~Freiwald}
\author[uniw]{C.~Hempfling}
\author[ethz]{D.~Hildebrand}
\author[ethz]{G.~Hughes}
\author[ethz]{W.~Lustermann}
\author[unige]{E.~Lyard\corref{cor1}}
\author[uniw]{K.~Mannheim}
\author[uniw]{K.~Meier}
\author[ethz]{S.~Mueller}
\author[ethz]{D.~Neise}
\author[unige]{A.~Neronov}
\author[tudo]{A.-K.~Overkemping}
\author[uniw]{A.~Paravac}
\author[ethz]{F.~Pauss}
\author[tudo]{W.~Rhode}
\author[uniw]{T.~Steinbring}
\author[tudo]{F.~Temme}
\author[tudo]{J.~Thaele}
\author[unige]{S.~Toscano}
\author[ethz]{P.~Vogler}
\author[unige]{R.~Walter}
\author[uniw]{A.~Wilbert}

\address[ethz]{ETH Zurich,~Institute for Particle Physics, Otto-Stern-Weg 5, 8093 Zurich, Switzerland}
\address[unige]{University of Geneva, ISDC Data Center for Astrophysics, Chemin d'Ecogia~16, 1290 Versoix, Switzerland} 
\address[uniw]{Universit\"at W\"urzburg, Institute for Theoretical Physics and Astrophysics, Emil-Fischer-Str.~31, 97074 W\"urzburg, Germany}
\scriptsize{\address[rwth]{Now at: RWTH Aachen, Physics Institute III A,  Sommerfeldstraße, 52074 Aachen, Germany} }
\address[tudo]{TU Dortmund, Experimental Physics 5, Otto-Hahn-Str.~4, 44221 Dortmund, Germany}

\cortext[cor1]{Corresponding authors}

\begin{abstract}	
The FACT telescope on the Canaries island of La Palma is the first
Imaging Atmospheric Cherenkov Telescope (IACT) to  use solid state
photomultipliers. It generates up to two terabytes  of data per night
which motivated us to investigate how to reduce  the volume of data.
Reducing the throughput enables us to efficiently  acquire, store and
process the observations data. This document  presents the conclusions
of this work, along with the implementation of the custom compression
algorithm and I/O layer that is currently in use to operate the
telescope. 

\end{abstract}


\end{frontmatter}

%\linenumbers

\nocite{*}

\section{Introduction}
The First Geiger-mode Avalanche photodiode (G-APD) Cherenkov Telescope
(FACT) has been operating on the Canary island of La Palma since October 2011.
 Operations were automated so that the system can be operated
remotely\footnote{http://www.fact-project.org}. Manual interaction is
 required only when the  observation schedule is modified due to weather 
conditions or in case of unexpected events such as a mechanical failure 
\cite{factDesign,factPerformance}. Automatic operations enabled high data taking 
efficiency, which resulted in up to two terabytes of FITS files \cite{fits} being recorded nightly and
transferred from La Palma to the FACT archive at ISDC in Switzerland.
Since long term storage of hundreds of terabytes of observations data is
costly, data compression is mandatory. This paper discusses the design
choices that were made to increase the compression ratio and speed of writing
 of the data with respect to existing compression algorithms.\\ 

Following a more detailed motivation, the FACT compression
algorithm along with the associated I/O layer is discussed. Eventually,
the performances of the algorithm is compared to other
approaches. 

\section{Motivations} 

%\begin{figure}[h!]
%\centering
%\includegraphics[width=0.95\textwidth]{IngestPipelines2.pdf}
%\caption{File transfer pipeline from data taking to off-site archive. The raw data is first written 
%to an on-site repository. FITS files are then compressed (gzipped) before being uploaded to a temporary
%off-site storage in Switzerland. The gzipped files are uncompressed so that they can be verified. 
%Upon success, gzipped files were moved to the long-term archive. The usage of compressed
%FITS allowed us to drop the compression steps (orange boxes).}
%\label{fig:transferPipeline}
%\end{figure}

A typical night of data taking generates up to two terabytes
of raw data. Several compression algorithms were considered and tested to
evaluate their performance on Cherenkov data. Lossy compression was
discarded to retain all bits of information coming from the detector.
As the first telescope using a new technology for photo detection, this
choice was important to avoid any bias in the characterization of the
G-APDs. Insufficient throughput disqualified the
slowest algorithms (lzma \cite{lzma} and bzip2 \cite{bzip2}) even though they
provided excellent compression. gzip \cite{gzip} was selected for the  
commissioning of the telescope.

%We selected gzipped \cite{gzip} FITS-files as
%a compromise between speed and efficiency during commissioning phase. The use of gzip allowed us to transfer the data
%to the swiss archive, but it introduced a compression step before sending the
%files through the Internet (figure \ref{fig:transferPipeline}). On the receiver side, incoming files
%must be uncompressed to a temporary location so that their integrity can be verified 
%before moving them to the long-term archive (\cite{ftools}).\\ 

The classical separation between raw file format and compression
introduces several reads and writes to the storage which could be
avoided if the file format would natively support compression. Several
formats such as HDF5 \cite{hdf5} and ROOT \cite{root} support
compression natively but are not widely employed by the astronomers
community. On the other hand, Tile-compressed FITS \cite{compressedFits1}
is a convention that allows to store compressed image data into binary
tables. A recent evolution of
the convention enables to also compress binary tables natively 
\cite{compressedFits2}. This convention is fully FITS compliant
with extra header keywords added to accommodate for the compression.

%Using this format allowed us to remove the extra compression step and continue
%to use our analysis pipeline without modification.\\ 

The current Tile-compressed FITS implementation is available from the CFITSIO library
and via a set of two executables, fpack and funpack, that can compress and decompress 
FITS files\footnote{http://heasarc.nasa.gov/fitsio/fpack}. Various compression algorithms 
are already included \cite{compressedFits}. Even though the compression of images was a fully 
functional feature, the handling of binary tables remained experimental at this time. 
Moreover, the specific noise pattern of the analog ring buffer (see figure \ref{fig:rawcalibPixel}) motivated
us from investigating a specific way to compress this data. Our primary goal was
to improve the compression ratio of raw data while maintaining a high throughput. 
We also wanted to explore the possibility to use calibration data of the analog ring buffer to increase
the compression ratio of raw data. To make sure that this calibration data
is exploited in the best possible way it had to be applied by the I/O layer, thus making
the compression process more complex.

%To allow for a fast and timely availability of the results of
%an analysis, the data must be read into the analysis chain once
%recorded. While writing and simultaneous reading already puts a high
%load on the disks, an additional intermediate compression step would
%not only delay the analysis but also significantly increase the load
%above a limit which can easily be maintained with standard systems today.
%Therefore, it is desirable to compress the data before it is written to 
%the disk to help reduce the load of the system even further.
%Eventually, the implemented compression algorithm could 
%improve compression ratios while increasing compression speed.

\section{Compression}
\label{sec:compression}
\begin{figure}[h!]
\centering
\includegraphics[width=0.45\textwidth]{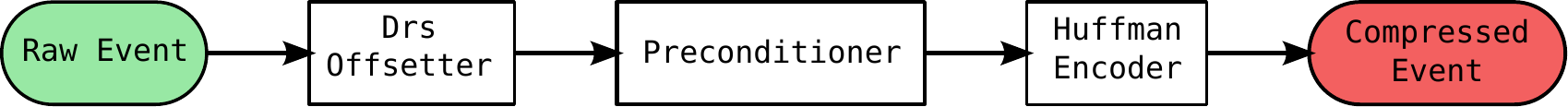}
\caption{Overview of the FACT compression I/O layer. Each Cherenkov event undergoes three processings. First the data
is calibrated (DRS). It is then preconditioned and eventually compressed using a Huffman encoder.}
\label{fig:compressionPipeline}
\end{figure}

An overview of the implemented I/O layer can be seen in figure
\ref{fig:compressionPipeline}. The FACT algorithm works on 
sixteen-bits integer data and is lossless
to allow to reconstruct the original raw data from the compressed output.
Our algorithm consists of three separate steps. 
First the data is reversibly pre-calibrated to
reduce the quasi random noise coming from the readout system to allow for 
high compression rates. Then the data is preconditioned
such that number of bits needed for storage is reduced. Eventually, 
the information is transformed such that only these significant bits are
stored. These steps are discussed in more details in sections \ref{sec:drscalib}, 
\ref{sec:preconditioner} and \ref{sec:huffman} respectively.

\subsection{Drs Calibration}
\label{sec:drscalib}

\begin{figure}[h!]
\centering

\begin{subfigure}[b]{\textwidth}
\includegraphics[width=0.25\textwidth]{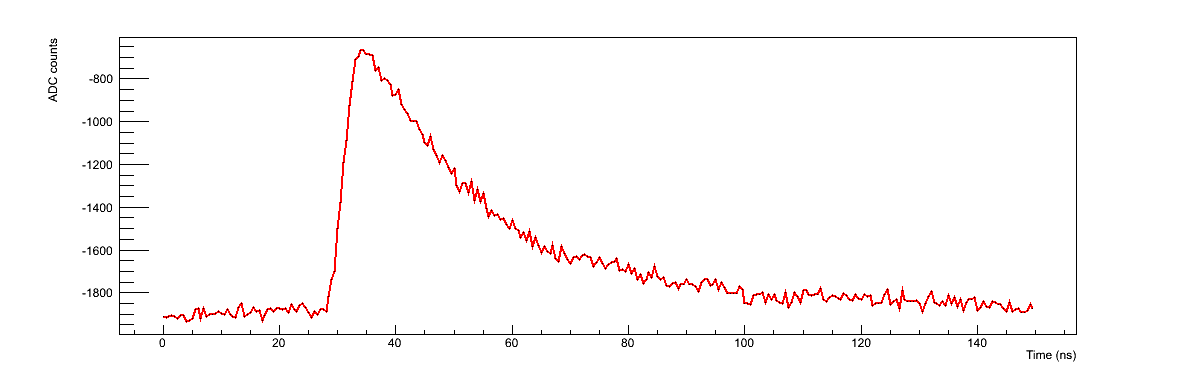}%{plots/pixel_curves_raw.png}
\includegraphics[width=0.25\textwidth]{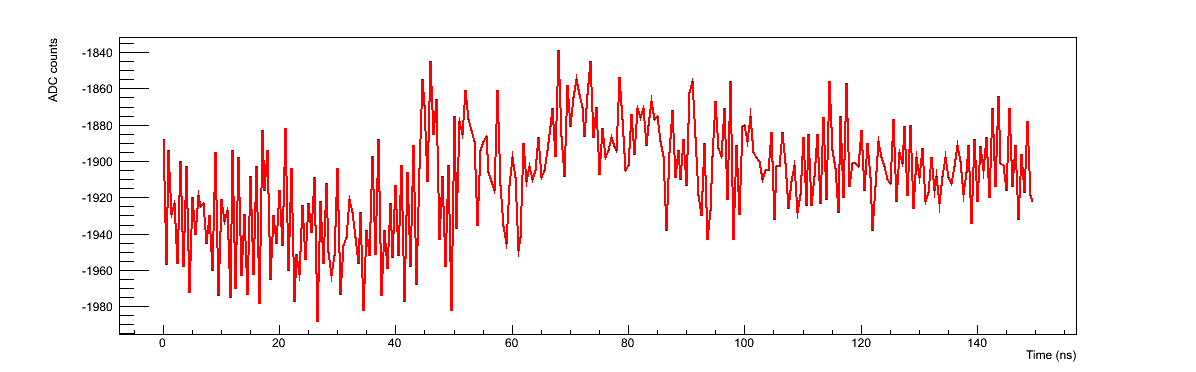}%{plots/pixel_curves_calib.png}
\end{subfigure}

\begin{subfigure}[b]{\textwidth}
\includegraphics[width=0.25\textwidth]{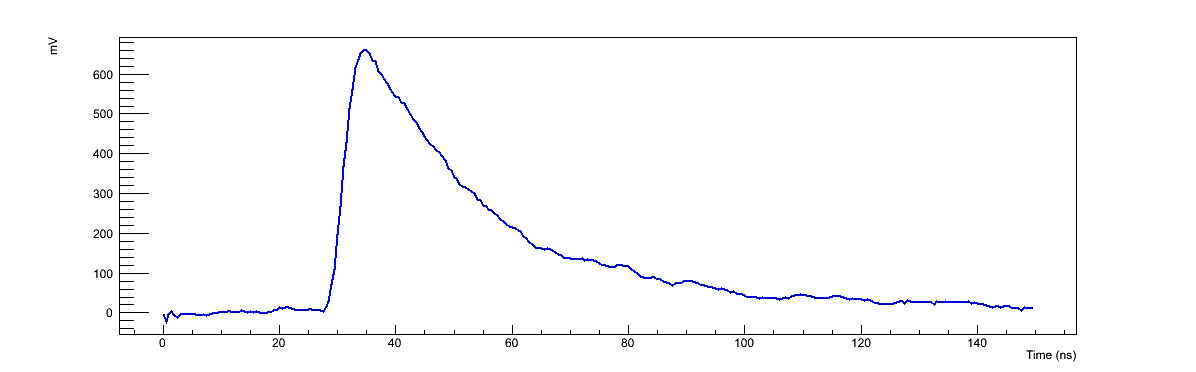}%{plots/pe_pixel_raw.png}
\includegraphics[width=0.25\textwidth]{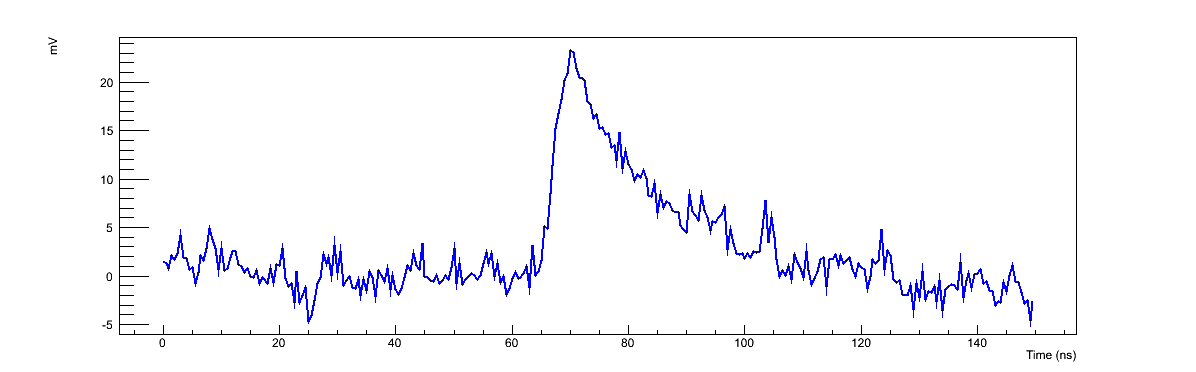}%{plots/pe_pixel_calib.png}
\end{subfigure}

\caption{Effect of DRS-calibration on a single pixel's waveform samples. 
Top, red: raw data coming from the DRS4. Bottom: DRS calibrated values. 
Left: large, 60 p.e. Cherenkov event. 
Right: small, 2 p.e. dark count event. The raw data outputted by the FACT camera is sampled by a 12-bits ADC
and stored using 16-bits integers. The actual DRS calibration transforms the samples into floating-point values, 
while the values are truncated to integers for compression purposes. }
\label{fig:rawcalibPixel}
\end{figure}

The FACT Camera \cite{factDesign} employs a Domino Ring Sampler chip (DRS\,4, \cite{ritt2008design})
to continuously record the signal from the detector plane. This chip is an analog ring buffer
that stores the signal before it is digitized and acquired. Each DRS4 sample
has an individual offset that must be calibrated using pedestal values. 
Upon trigger, the sampling is halted and the content of the buffer is
readout. A trigger can happen at any time, thus the  readout window is
different from event to event. Therefore, not only the offsets are
individual to each sample but also vary from event to event.
Although strictly speaking the offsets that are added to the raw signal 
are deterministic, they will appear as random to most algorithms and significantly
decrease the achieved compression ratio.
To compensate for this effect, the position of the
readout window is recorded as well so that the raw data can be
calibrated with the offsets individual to each sample
before being analyzed  \cite{DrsCalib}. In Figure \ref{fig:rawcalibPixel} (top) 
a raw signal is compared with a calibrated one (bottom)
while the statistics for an entire event can be seen on figure 
\ref{fig:preconditionning}.

Applying this calibration reduces the amount of pseudo-random noise. 
The downside is that this moves the data from integer
to floating-point space which would introduce additional bytes in each
sample. To stay in integer space and allow to reverse the process, the compression algorithm only applies 
the integer part of the offset. Since the majority of offsets is larger than unity,
this does not significantly change the width of the resulting
distribution (by only 0.0011). Another advantage is that for a first look at the 
data this calibration is fairly sufficient and, therefore, allows
easy access to semi-calibrated data. The offsets themselves are stored in each data file thus
allowing to reverse this simplified calibration and apply a more precise one. 
This is entirely done on-the-fly by the I/O layer in a transparent way for the user who is 
only aware of the raw values.

%better compression as it reduces the level of noise in the dataset.
%Figure \ref{fig:preconditionning} shows that calibrating the pixel
%values  reduced the total number of values, hence better compression
%ratios.

\subsection{Preconditioner}
\label{sec:preconditioner}

While the DRS calibration step reduced the amount of noise,
 the data still contains a significant
amount of pulses originating from so-called dark noise which require a
comparably high number of bits for storage. Since these pulses look just
like photons signal and thus have slow slopes compared to the sampling 
frequency, storing their derivative instead of their 
amplitude should reduce the required number of bits. 
If just the difference between two 
consecutive samples would be recorded, the noise
from two samples would sum up yielding increased random noise 
on the differences. The comparably slow change of the signal, however,
allows us to average two consecutive samples and use them to calculate
the difference to the following one. In this way the additionally
introduced noise component is reduced by \(\sqrt{2}\). Averaging
a higher number of samples does not significantly lower the
additional noise, but becomes sensitive to faster signals. Therefore,
the average of two has been found to be the best compromise.

The preconditioned samples \(p_i\) are calculated from
the DRS calibrated raw data samples \(s_i\) by
\[p_i = s_i - \frac{s_{i-1}+s_{i-2}}{2}\]
with
\[p_0 = s_0\qquad\mbox{and}\qquad p_1 = s_1\]

%\begin{algorithm}
%\caption{Preconditioner for ADC values}
%\label{algo:precond}
%\begin{algorithmic}
%\REQUIRE $s$ an array of integers of size $n$
%\FOR{$i=n-1$ \TO $2$}
%\STATE $s[i] \leftarrow s[i] - \frac{s[i-1]+s[i-2]}{2}$
%\ENDFOR
%\end{algorithmic}
%\end{algorithm}

\begin{figure}[h!]
\centering
\includegraphics[width={0.4\textwidth}]{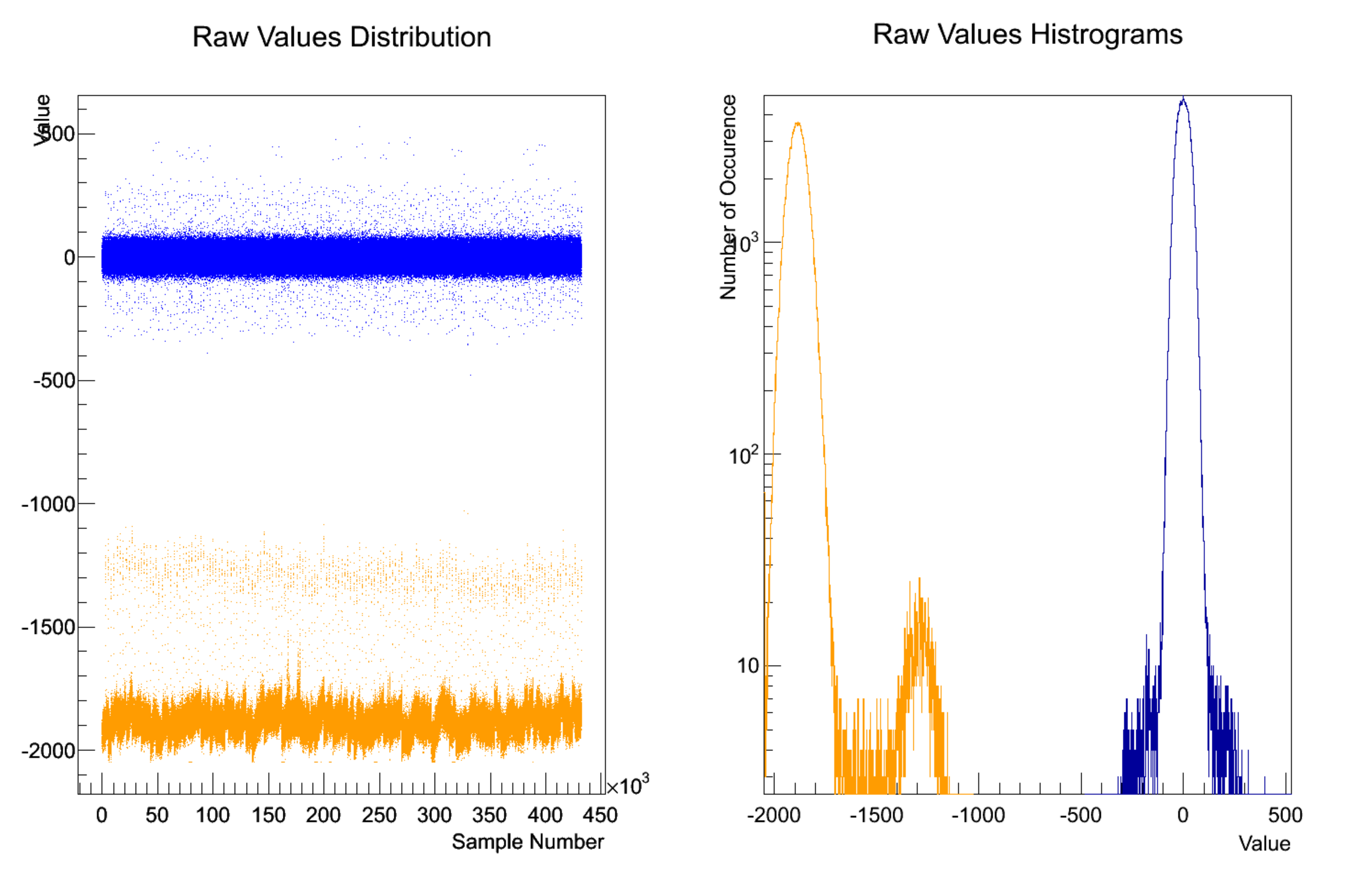}
\includegraphics[width={0.4\textwidth}]{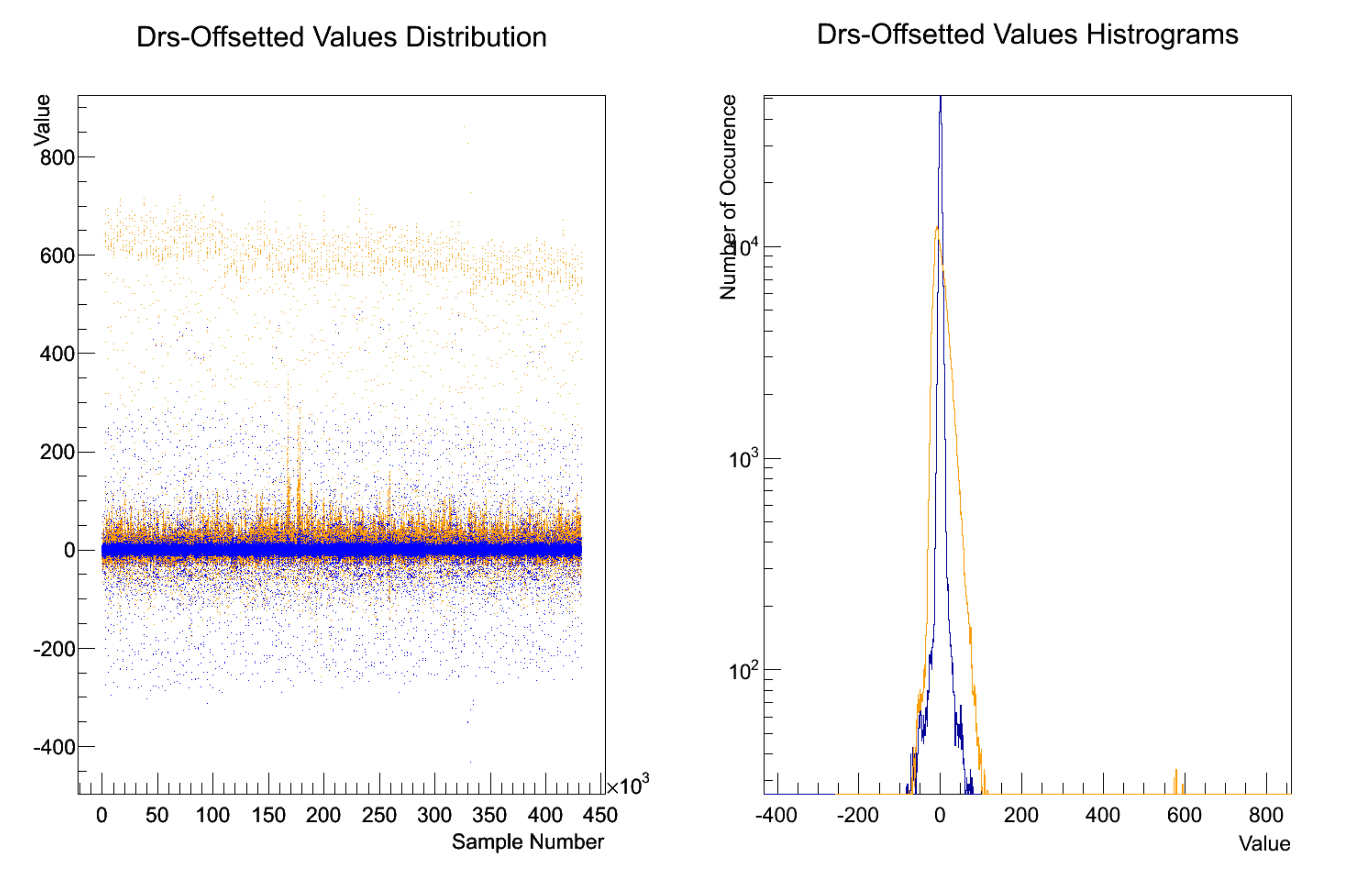}
\caption{Plotting of the pixel values from a single event. Each event has 1440 pixels and for each pixel 
300 time samples of 0.5ns are recorded. On top are the values distribution and histograms for raw data, while
below are the same plots for DRS-calibrated data. In orange are the input data while in blue are the
output values of the preconditioning.
The mean and std. dev. are (-1881.95, 65.0274), (-0.25, 35.63), (5.53, 49.09) 
and (-0.01, 15.21) for raw, preconditioned raw, DRS-calibrated and 
preconditioned DRS-calibrated respectively.
The second peak in the histogram of the raw data comes from artificially added time markers employed to 
synchronize the trigger patches.}
\label{fig:preconditionning}
\end{figure}

In practice, this algorithm reduces the number of different values,
while the occurrence of values close to zero is increased (Figure \ref{fig:preconditionning},
bottom). The width of the signal is reduced from 65 to 35 and from 49 to
 15 counts for raw and DRS-calibrated data respectively.

The original waveform is restored by
\[s_i = p_i + \frac{p_{i-1}+p_{i-2}}{2}\]
with
\[s_1 = p_1 \qquad\mbox{and}\qquad s_0=p_0\]

%\begin{algorithm}
%\caption{Recovery of precontioned ADC values}
%\label{algo:decond}
%\begin{algorithmic}
%\REQUIRE $val$ an array of integers of size $n$
%\FOR{$i=2$ \TO $n-1$}
%\STATE $s[i] \leftarrow s[i] + \frac{s[i-1]+s[i-2]}{2}$
%\ENDFOR
%\end{algorithmic}
%\end{algorithm}

\subsection{Huffman Encoding}
\label{sec:huffman}

The last step of our compression algorithm consists of minimizing 
the number of bits needed to store the data. This is achieved via entropy coding
and more specifically using Huffman coding \cite{huffman}. The
principle of this encoding is to associate codes to symbols, which lengths are inversely
proportional to the symbols occurrence count. It has been widely used
over the past 50 years, with well known applications such as JPEG\footnote{http://www.itu.int/rec/T-REC-T.81-199209-I/en} and MP3\footnote{http://mpeg.chiariglione.org/standards/mpeg-1/audio}.

The length of the input symbols can vary depending on specific
quantization parameters, especially for lossy compression.
8-bits words is the most commonly used symbol length for lossless
compression. We chose 16-bits words as
input to overcome the necessity to code the
zeros in the majority of the most significant bytes of the 16-bit
data. 
%Reducing the length of the input words exactly to 12
%bits would have had no effect on the final compression ratios because
%Huffman discards unused input symbols by design.

An alternative approach to overcome this shortcoming is the Rice
algorithm \cite{rice}, which separates the high and low parts of
the words before compressing them separately.

\section{File format}
\label{sec:fileformat}
As a file format, the FITS tile-compression convention has been chosen. The use of
Tile-compressed FITS \cite{compressedFits1} allowed us to reuse most of the
analysis pipeline and FTOOLS.
For instance, all data files are verified by the FTOOL \emph{fitsverify} before 
being accepted to the long-term archive. Our own FITS I/O layer had already been 
developped within the FACT project for performances reasons and more particularily
to be able to control the memory allocations. These classes were 
extended to accomodate for the compression algorithms and Tile-compression
convention. To implement the compressed format it was enough to derive from the
existing classes, thus reducing the required development work to a minimum.\\

FITS-files are organized in \emph{extensions}. Each extension starts with an 
ASCII header that defines the structure and length of the data stored in
the current extension. Headers are directly followed by fixed-length data organized
into columns and eventually comes the variable-length data in a section called the \emph{heap}.  
The heap contains compressed data that could not be stored
in the fixed-length columns of the main data table. In the compressed format, 
rows are compressed in groups of $n$ (called a \emph{tile}) and each column from a group of rows 
is compressed separately, as directed by the Tile-compressed FITS convention. \\

\begin{figure}[h!]
\centering
\includegraphics[width={0.3\textwidth}]{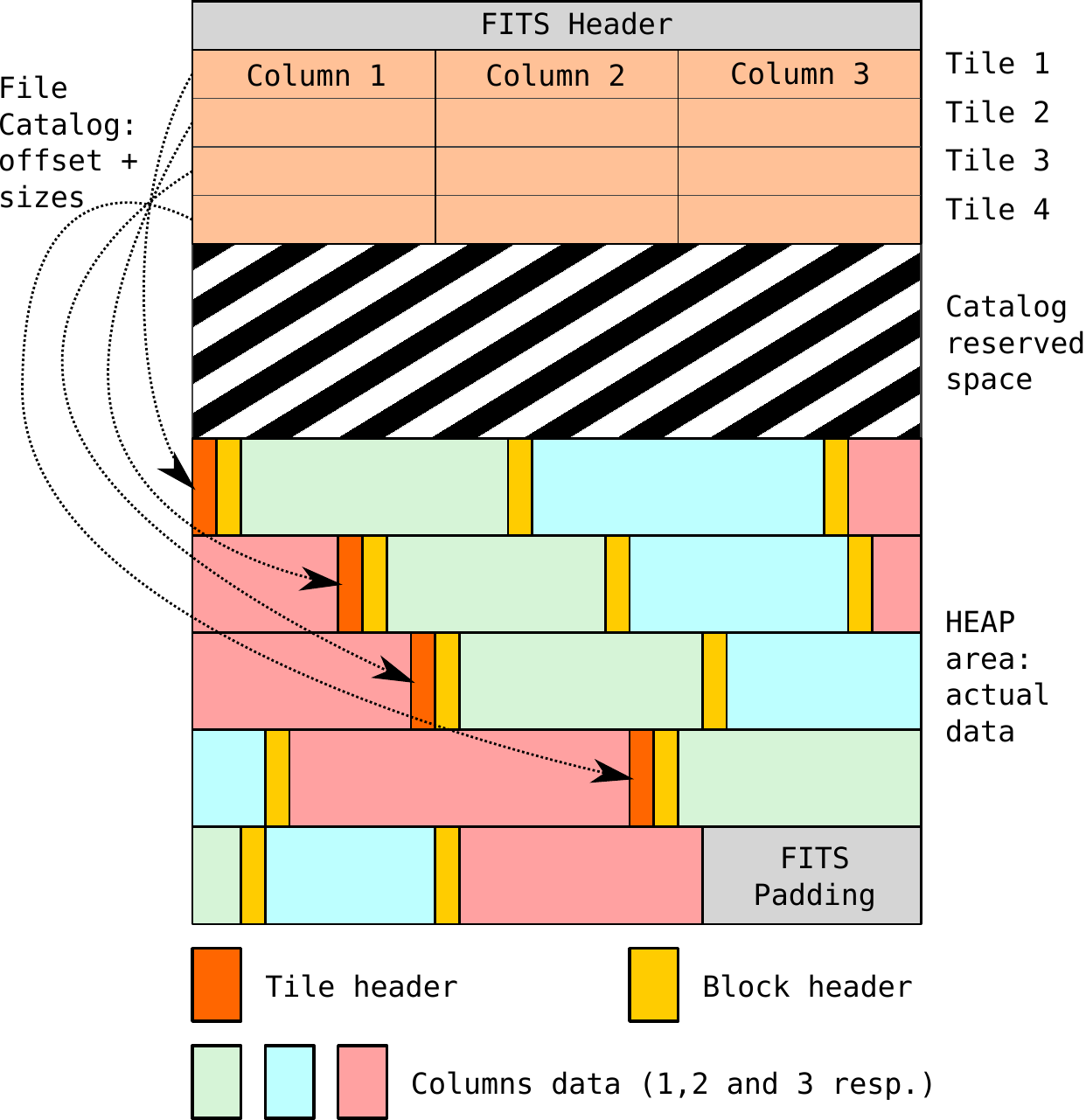}
\caption{Schematic view of the data layout on disk. The file format complies with the 
Tile-compressed FITS convention. All columns, even uncompressed ones are moved to the heap area for
better performances. The main data table thus becomes a catalog. The tile and block headers 
allow to reconstruct the file's catalog from the heap only. This is used in emergency
situations where not enough space was reserved. It could also be useful if each block of 
events (tiles) are stored in relational databases (RDMS). This way complete FITS files could
be efficiently reconstructed on the fly based on the request made to the RDMS.}
\label{fig:fileformat}
\end{figure}

A schematic view of the file format can be 
seen in figure \ref{fig:fileformat}.
In Tile-compressed FITS files, all compressed data goes into the heap 
as it is naturally variable in length. In our implementation we also allow 
for the storage of fixed-length data in the heap. This allows to read 
continuous sectors on the disk to retrieve a full set of rows, rather 
than alternating reads between the columns and heap areas. In this way 
the streaming capability of the format is kept. Moreover, storing all 
data to the heap simplifies the structure of the I/O code, thus easing 
the long-term maintenance. \\

A few minor adaptation were implemented to address some shortcomings of
the Tile-compressed FITS format when working with streams rather than data
sets. These change do not modify the capabilities of FITS but rather are meant
to make operations more robust and less resources intensive.
They are discussed in details in the following sections. 

\subsection{Custom FITS keywords}
Two new header keywords were introduced: RAWSUM and ZSHRINK. 

\paragraph{RAWSUM} In the standard implementation of the Tile-compressed FITS
format, the checksum of the uncompressed table (DATASUM) is saved during
compression by the fpack tool.

While in the FITS convention this checksum is calculated
from big-endian data, the data arriving from the telescope is
little-endian. To avoid the need of an additional byte swapping
just to calculate the DATASUM, a new keyword RAWSUM has been introduced
storing the checksum of the uncompressed, little-endian data. 
Appart from the omission of the byte-swap, the computation of both
DATASUM and RAWSUM is identical. This new keyword does not 
forbid to use the usual CHECKSUM keywork to verify the integrity of 
the compressed data, which we do.

%The reason behind the introduction of RAWSUM stems from the current
%implementation of the FITS compressor, called fpack.  It is capable of
%transcoding existing FITS files, but not to compress a stream of data.
%Hence, if there are checksums in the input FITS files, they are saved in
%new keywords while the checksum of the compressed tables replace the
%existing keywords. In our case, there is no checksum to translate
%because the data is written to compressed  FITS directly from the
%stream coming from the telescope. Thus the original datasum is computed
%on-the-fly with big-endian data. This datasum is not the same as the
%regular FITS datasum which is computed on little-endian data, hence the
%new keyword to bypass a systematic bytes-swapping.\\

\paragraph{ZSHRINK}

In our Tile-compressed FITS streamer the main data table is used to store 
pointers to the heap-area, usually one for each compressed row. 
Since this table is stored before the heap-area, it implies that the number of rows to 
be written to the file is known in advance. 
This is not always possible because experiments often prefer to group data per 
interval of time rather than per unit of raw data.

As a remedy, an estimate of the expected number of rows/events
is calculated when the file is opened, and a corresponding
number of bytes reserved. If the number of events grows larger
than the number of pointers which can be written to the table,
only every $N$th pointer is written to the table. The integer 
\(N\) is then stored in the keyword ZSHRINK.

To still be able to read the compressed data entirely, specific
markers are put in the heap area that allow us to not only read the
compressed data without access to the main data table, but also to
reconstruct this table entirely from the heap area only. 

\subsection{Compression blocks}

\begin{figure}[h!]
\centering
\includegraphics[width={0.2\textwidth}]{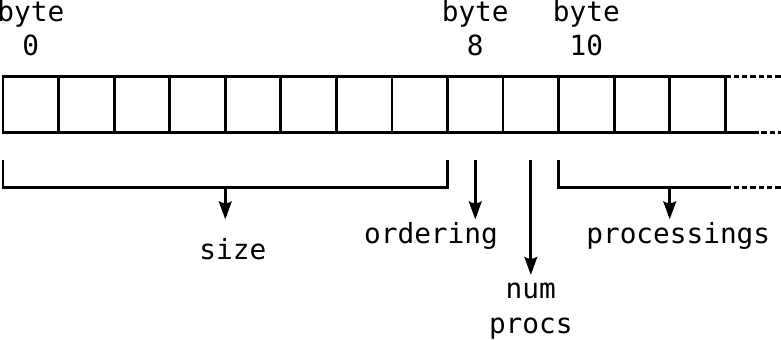}
\caption{Compressed block header. $size$ (8 bytes) is the size in bytes of the 
block, $ordering$ (1 byte) is how the data was copied from the columns. ASCII code
for 'R' means that the data was copied as is, while 'C' means the it was transposed as 
directed by the Tile-compressed FITS convention. 
$num procs$ (1 byte) is the number of compression algorithms that were applied and
$processings$ ($num procs$ bytes) is the identifiers of the applied algorithms. 
Currently the only valid values are $0=raw$, $1=smoothing$ and $2=Huffman16$.}
\label{fig:blockHeader}
\end{figure}

Inside the heap, blocks of compressed 
data start with their own variable-length header which is described in
figure \ref{fig:blockHeader}. An $ordering$ field allows us to order the
data either by row or by column. In the case of FACT, ordering the data
per row yields a better compression ratio and less data shuffling.

Besides block headers, tile headers are interleaved with the compressed
data. If the main data table is complete, i.e. contains pointers to
all rows, these headers are redundant. If more rows are written
compared to the space reserved initially (ZSHIRNK$>1$), they allow to 
reconstruct the missing main data table entries. They can also be used
to recover the catalog upon data corruption.

\begin{figure}[h!]
\centering
\includegraphics[width={0.2\textwidth}]{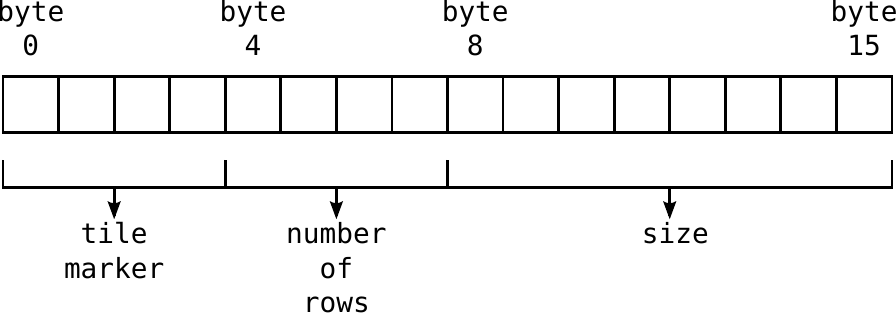}
\caption{Tile header. $tile marker$ is an identifier marking the start of a new tile. 
Its content is the ASCII codes of $TILE$. $number of rows$ is the number of rows compressed
in the current tile. $size$ is the total size of the tile in bytes. }
\label{fig:tileHeader}
\end{figure}

\subsection{DRS-calibration}
The DRS-calibration is not applied by the compression algorithm per se.
The calibration  offsets are stored in a separate Tile-compressed table in
the same FITS file as the data itself.  This introduces an overhead
which is quickly absorbed by the increased compression  ratio. For
optimal efficiency, the calibration table is  always placed before the
main data table. This allows the data access layer to quickly  find the
calibration table without having to skip through the file. 
The data access layer finds the calibration table by looking for a
table called  \emph{ZDrsCellOffsets}, and thus this name should not be
used for another table. 

\subsection{Access Layer}
The access layer was written in C++ and consists of classes that
inherit from each other, as follow: \emph{factfits} $\rightarrow$
\emph{zfits} $\rightarrow$ \emph{fits}. Additionally, \emph{fits} has
the zlib as a link option which enables it to read gzipped FITS
natively. \emph{zfits} can read everything \emph{fits} can plus 
Tile-compressed FITS and \emph{factsfits} can read all of what \emph{zfits} can 
plus DRS-calibrated FITS. The source code of the classes can be checked-out
from the FACT svn\footnote{https://www.fact-project.org/svn/trunk/FACT++}. All reading classes are
single-threaded while their writing counterparts use multiple cores for
fast compression.

The writing is done as follow: incoming rows are buffered until the
target number of rows per  tiles is reached. Then the buffer is given
to a compression thread while a new one is allocated to receive new
events. Each compression thread will shuffle and compress the data
before passing the compressed rows to a thread that perform the actual
write to disk.

Reading occurs as follow: compressed tiles are loaded to
memory, decompressed and the raw data is buffered until it is requested
by the users or until another tile is read.

Our I/O layer allows to start reading a file before it has been closed by the writing
process. This proved to be useful to start real-time analysis as soon as possible
and to provide statistics to the operator if they need to be extracted from raw data files. 

\section{Results}
\label{sec:results}
The FACT compression algorithm was compared to other de facto
standards in the astrophysics community. Since the DRS calibration is
specific to the DRS\,4 readout, two separate input data sets
were employed: raw events and their DRS-calibrated version. The
calibration step was pre-calculated for all tests so that the
comparison is fair. The tested file formats were:

\begin{itemize}
\item \emph{EVENTIO} is the format used to produce Monte-Carlo simulations for the CTA project\footnote{http://www.mpi-hd.mpg.de/hfm/~bernlohr/iact-atmo/}. This format tightly packages the input data
using a internal compression algorithm. 
\item \emph{FITS} the reference file format where the data is written in plain binary. Only the bytes order might be modified to make the data big-endian.
\item \emph{Tile-compressed FITS} the FITS convention that supports binary tables compression
natively.
\end{itemize}

In addition, HDF5 and ROOT files were considered but not tested because
the way they package the data overlapps with the above formats. HDF5 defines
the file format but not the compression algorithm to be used per-se, while ROOT
uses a derivative of gzip.
FITS was used as a data source format, while Tile-compressed FITS was split into
two flavors:

\begin{itemize}
\item \emph{Rice}: the FTOOLS fpack and funpack were used to (un)compress the data using the
Rice compression algorithm \cite{rice}. This algorithm splits high and low bytes of the input values
before compressing them separately.
\item \emph{Fact}: the compression algorithm and I/O layer described in this paper were used to produce
Tile-compressed FITS. 
\end{itemize}

Besides file formats, each produced file has been further compressed
using several classical algorithms, namely:

\begin{itemize}
\item \emph{lzma} the Lempel-Ziv-Markov chain-Algorithm that was under development until 2001. 
\item \emph{gzip} the well known algorithm widely used by the linux community. It employes the LZ77 and Huffman coding
\item \emph{bzip2} the more recent algorithm meant as an alternative to gzip. It uses the Burrows-Wheeler transform and Huffman coding.
\end{itemize}

All programs that have compression level options were set to their
minimum. Preliminary tests conducted with higher levels showed that
the achieved ratios did not increase significantly whereas the
processing time did.

Tests were conducted on data from 2 nights: 2014/01/01 and 2014/01/10. 
The first night was dark, while the second had moonlight for half of
the night resulting in more recorded background photons and thus a 
higher noise level. All
I/O operations were done to/from the shared memory to alleviate caching
effects  and storage bottlenecks. Only files small enough to fit in
the available memory were processed\footnote{71 out of 261 
runs could not fit in the available 64GBytes of memory}. Since only the FACT I/O layer
 could handle the  DRS-calibration natively, the input data
sets were pre-processed before the other compression algorithms were applied. The size of 
the DRS-calibration table was ignored in the calculations. This table has a constant size of 
approximately 1.4MB compressed or 0.02 percent of the average raw data file, to be added to 
the output size of the compressed files. The raw throughput of the memory was approx. 1.5GB/s 
thus the performances given below reflect the computation time rather than memory IO.

\begin{figure}[h!]
\centering
\includegraphics[width={0.45\textwidth}]{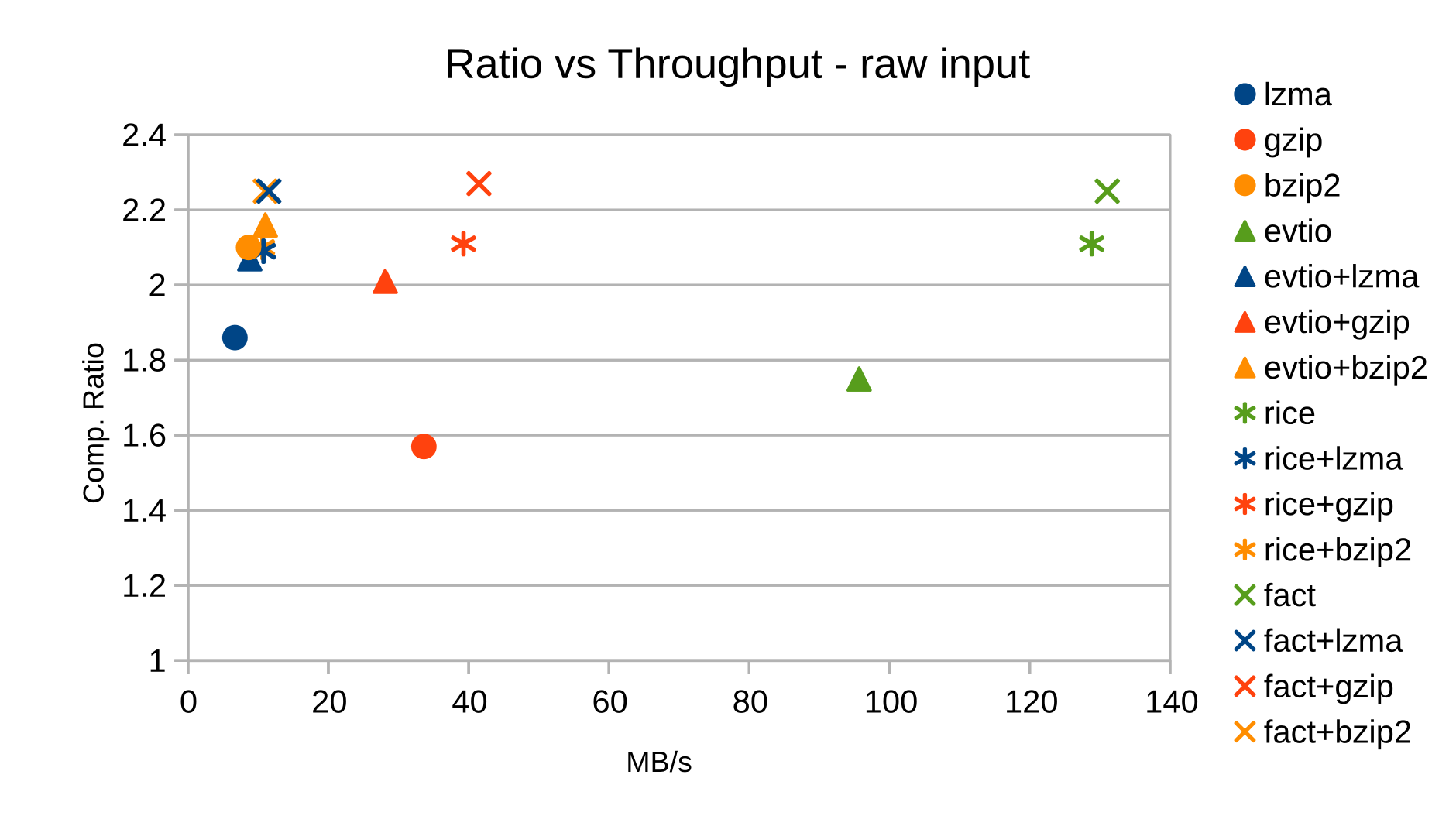}
\caption{Obtained compression ratio compared to throughput for raw data set. The best performances are obtained by the data points with the largest x and y values.}
\label{fig:summaryRaw}
\end{figure}

\begin{figure}[h!]
\centering
\includegraphics[width={0.45\textwidth}]{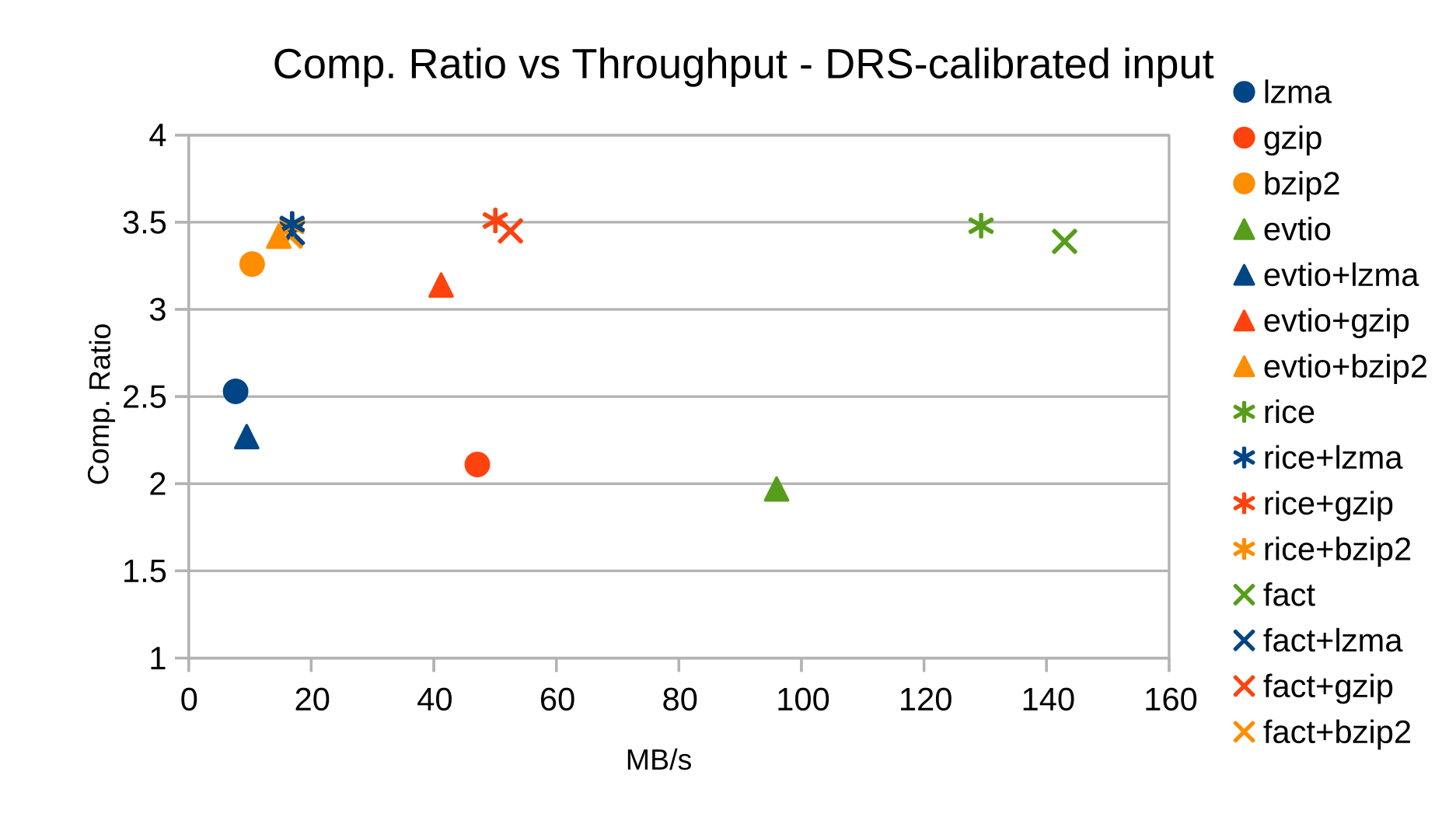}
\caption{Obtained compression ratio compared to throughput for DRS-calibrated data sets.}
\label{fig:summaryCalib}
\end{figure}

A summary of the compression performances can be seen on figures
\ref{fig:summaryRaw} and \ref{fig:summaryCalib}. The compression ratio
for the raw data went up to 2.25, while the I/O layer delivers the 
best average ratio and an average throughput of 131MB/s. 
For Drs-calibrated data, we achieved ratios up to 3.5 with 
Rice. The FACT algorithm is a close contender, 
achieving an average ratio of 3.39 coupled with a throughput of 143MB/s.  

Decompression speed for calibrated data can be seen on table \ref{tab:meandecompcalib}. 
EventIO turned out to be the fastest format
when it comes to reading the data back, topping out at 200\,MB/s. 
The FACT algorithm was second at 132\,MB/s and Rice arrived third at
115\,MB/s. Classical algorithms performed much worse with only gzip coming
close to the Rice decompression performances. Detailed performances, 
including decompression of raw data, are given in the annex \ref{appendix:detailedResults}. 

\begin{table}
\begin{tabular}[h]{|c|c|c|c|c|}
\hline
  & FITS & EventIO & Rice & FACT \\
\hline
native & n/a &  200.51 & 109.29 & 120.34 \\
\hline
lzma   & 30.75 & 34.16 & 31.86 & 32.13 \\
\hline
gzip   & 94.14 & 86.64 & 83.92 & 90.57 \\
\hline
bzip2  & 26.80 & 28.51 & 34.05 & 34.77\\
\hline
\end{tabular}
\caption{Mean decompression throughput in MB/s of input DRS-calibrated data.}
\label{tab:meandecompcalib}
\end{table} 

\section{Discussion}
The tests have shown that the described compression algorithm provides
good performances when applied to the data produced by the FACT
telescope. Compared to the previously used gzipped-FITS format, it  allowed the experiment to 
improve the compression ratio of its raw data from  1.6 to 3.4 while the compression throughput
went from 33.6\,MB/s up to 143.0\,MB/s. Only Rice was able to outperform
our algorithm under specific conditions, namely for
DRS-calibrated data when the lighting conditions are not optimal. 
These degraded lighting conditions occur between the nautical and
astronomical twilight, and when the moon is up. This suggests that
the Rice algorithm is very good at compressing actual signal while 
our algorithm achieves larger compression ratios with raw data. 

Adding classical compression on top of the custom ones did not
significantly  improve the results and even decreased the compression
ratio in some cases. This was expected as the first compression stage
leave mostly noise in the data set.

The FACT algorithm is the fastet of the tested
 approaches, while Rice is second. The classical
algorithms were much slower and would be difficult to use for real-time
operations as they would require that the data is compressed after data taking
in a separate step or using a much larger number of compute cores. The throughput obtained by these algorithms is faster in some
cases where the data was first transcoded to a natively  compressed
format.

The best overall compression ratios were obtained by combining a native
algorithm - either Rice or Fact - and gzip. However, the gain in
compression ratio of about 1\% is not significant enough to
accept a decrease in processing speed of more than 50\%.

Considering the good performances of the Rice algorithm under poor
lighting conditions for DRS-calibrated data, an additional gain
could be achieved. However, given the marginal improvements
compared to our algorithm, that only a small fraction of the
data is taken under these conditions and the additional complexity to
apply two separate compressions, the FACT algorithm remains a good
choice.  

\section{Conclusion and Future Work}
In this paper a simple compression algorithm was presented which provides
good performances when applied to FACT data. This algorithm was
implemented in C++ and integrated into a streaming I/O layer that produces
Tile-compressed FITS file format, which makes it suitable for the real time operations of IACTs.

The performances of our algorithm was compared with existing
approaches. The experience gained during this work will be reused while
devising the raw data format for the Cherenkov Telescope Array to
ensure that the best compression ratios achievable in real-time is implemented.

The I/O layer described in this paper has been used since more than three years for the datataking
of FACT. The total amount of encoded data is currently more than 600TB uncompressed
and all raw data can be read without any problem.   

\section{Acknowledgments}
This work was made possible thanks to SNF Synergia grant, 
ETH Zurich grant ETH-10.08-2 as well as the funding by the
German BMBF (Verbundforschung Astro- und  Astroteilchenphysik). We are
thankful for the very valuable contributions from E. Lorenz,  D. Renker
and G. Viertel during the early phase of the project. We thank the
Instituto de Astrofisica de Canarias allowing us to operate the
telescope at the Observatorio Roque de los Muchachos in La Palma, and
the Max-Planck-Institut f\"{u}r Physik for providing us with the mount of
the former HEGRA CT3 telescope. We thank William Pence and Rob Seaman
for the time they spent providing guidance for the fpack  software and
their valuable discussion insights. We also thank Konrad Bernl\"{o}hr for
his EventIO example for writing FACT data.

%\section*{References}

\bibliography{bibliography}

\begin{appendices}

\section{Detailed results}
\label{appendix:detailedResults}
The detailed results are presented below. First the compression ratios and then the compression throughput. 
One plot with the tested formats is presented per classical compression ratio. In the case of the raw file 
format, plain FITS was omitted as the corresponding compression ratio is always one and calculating a throughput
would make no sense.

For both the compression ratios and throughput, the data points are organized per run. Each run corresponds
to a single raw data file that was moved to shared memory in plain FITS. All the code was compiled with 
gcc 4.4.7 on scientific linux 6.2 x64. The optimizer was set to -O2. Tests were made with -O3, but the lack 
of performances improvements made us stay with the -O2 option. 

The first half of each plot correspond to the night of 2014/01/01 (dark) up to run 78 while the second half corresponds to the night of 
2014/01/10 (moon). Jumps in the compression ratios correspond to repointings of the telescope, while jumps in the 
throughput are most likely due to system interrupts of the operating system of the server onto which the tests were run. Indeed, despite
the fact that we made sure that no other processing was running on the servers for the tests, some system interrupts
still occurred. The total input size was 1330.56GB while the average file size was 6.97GB.
 
Tables \ref{tab:meanratioraw} and \ref{tab:meanratiocalib} show the average compression ratios for
the raw and DRS-calibrated input data sets respectively.
A summary of the compression speeds can be seen on table \ref{tab:meanthroughputraw} and \ref{tab:meanthroughputcalib}. Eventually, 
the decompression speed can be seen on tables 
\ref{tab:meandecompraw} and \ref{tab:meandecompcalib}. 

\begin{table}
\begin{tabular}[h]{|c|c|c|c|c|}
\hline
       & FITS & EventIO & Rice & FACT \\
\hline
native & 1    & 1.75 & 2.11 & 2.25 \\
\hline
lzma  & 1.86 & 2.07 & 2.09 & 2.25 \\
\hline
gzip  & 1.57 & 2.01 & 2.11 & 2.27 \\
\hline
bzip2 & 2.10 & 2.16 & 2.10 & 2.25\\
\hline
\end{tabular}
\caption{Mean compression ratio for raw input data.}
\label{tab:meanratioraw}
\end{table} 

\begin{table}
\begin{tabular}[h]{|c|c|c|c|c|}
\hline
  & FITS & EventIO & Rice & FACT \\
\hline
native & 1    & 1.97 & 3.48 & 3.39 \\
\hline
lzma   & 2.53 & 2.27 & 3.49 & 3.44 \\
\hline
gzip   & 2.11 & 3.14 & 3.51 & 3.45 \\
\hline
bzip2  & 3.26 & 3.42 & 3.48 & 3.42\\
\hline
\end{tabular}
\caption{Mean compression ratio for DRS-calibrated data.}
\label{tab:meanratiocalib}
\end{table}

\begin{table}
\begin{tabular}[h]{|c|c|c|c|c|}
\hline
  & FITS & EventIO & Rice & FACT \\
\hline
native & n/a   & 95.68 & 128.84 & 131.05 \\
\hline
lzma   & 6.68  & 8.80  & 10.75  & 11.52 \\
\hline
gzip   & 33.62 & 28.11 & 39.27  & 41.47 \\
\hline
bzip2  & 8.62  & 11.01 & 10.61  & 11.02\\
\hline
\end{tabular}
\caption{Mean compression throughput in MB/s of input raw data.}
\label{tab:meanthroughputraw}
\end{table} 

\begin{table}
\begin{tabular}[h]{|c|c|c|c|c|}
\hline
  & FITS & EventIO & Rice & FACT \\
\hline
native & n/a &  95.97 & 129.33 & 142.96 \\
\hline
lzma   & 7.67 & 9.47 & 16.89 & 16.94 \\
\hline
gzip   & 47.11 & 41.21 & 50.05 & 52.52 \\
\hline
bzip2  & 10.36 & 14.70 & 16.89 & 16.59\\
\hline
\end{tabular}
\caption{Mean compression throughput in MB/s of input DRS-calibrated data.}
\label{tab:meanthroughputcalib}
\end{table} 

\begin{table}
\begin{tabular}[h]{|c|c|c|c|c|}
\hline
       & FITS & EventIO & Rice & FACT \\
\hline
native & n/a   & 193.63 & 114.87 & 131.90 \\
\hline
lzma   & 21.10  & 22.51  & 21.66  & 23.65 \\
\hline
gzip   & 79.55 & 82.58 & 85.75  & 95.53 \\
\hline
bzip2  & 23.72  & 24.10 & 23.99  & 26.07\\
\hline
\end{tabular}
\caption{Mean decompression throughput in MB/s of input raw data.}
\label{tab:meandecompraw}
\end{table}

\subsection{Compression Ratios}

\begin{figure}[p]
\centering
\includegraphics[width={0.45\textwidth}]{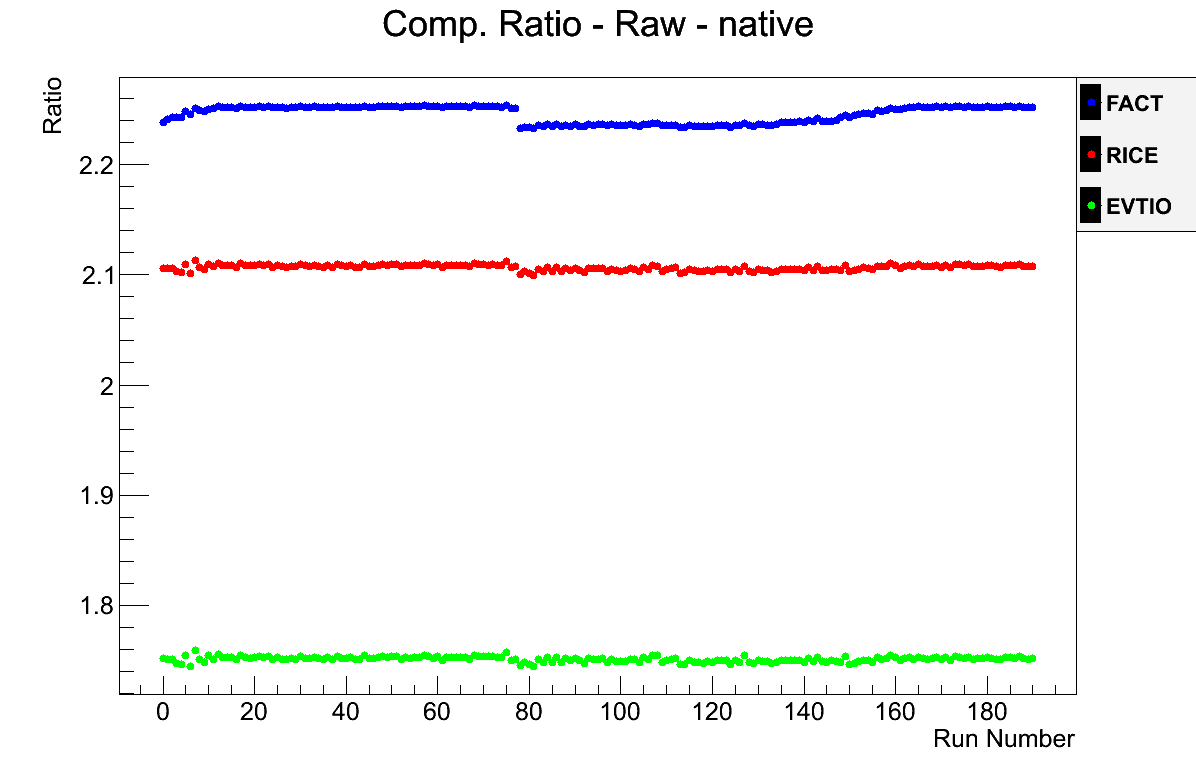}
\includegraphics[width={0.45\textwidth}]{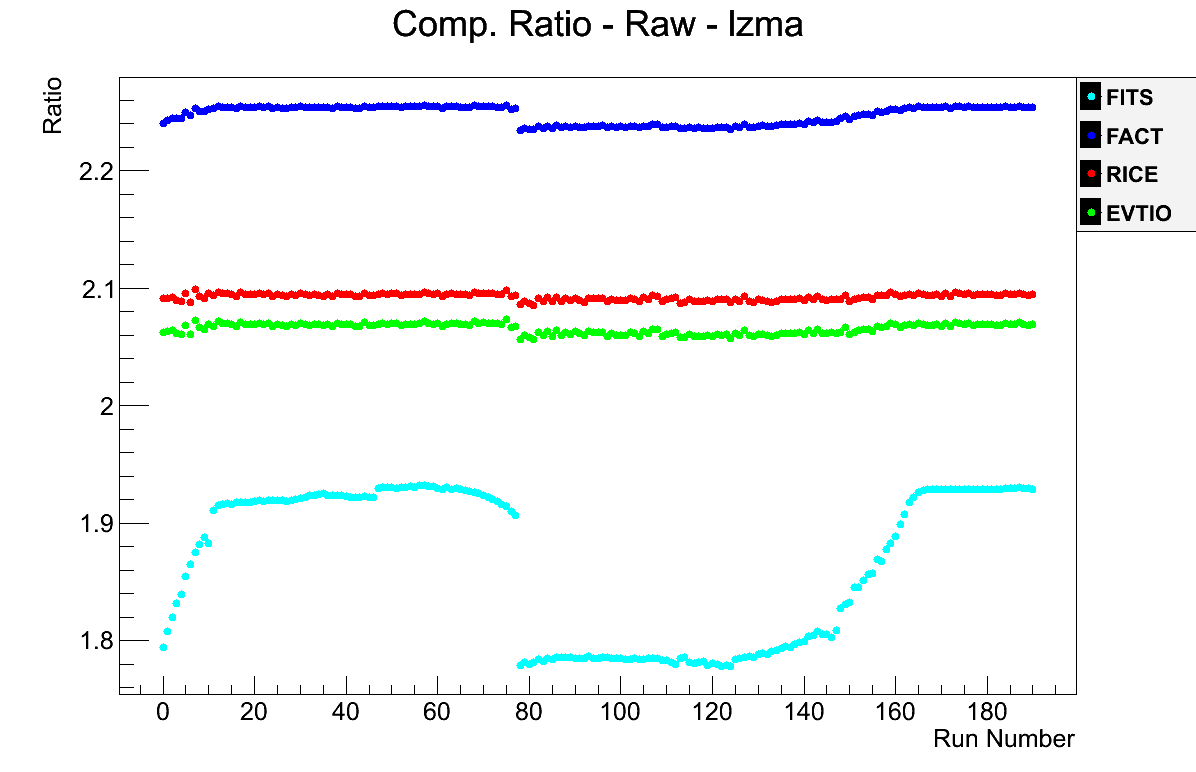}
\caption{Compression ratios of raw data runs for native and lzma output respectively.}
\label{fig:ratiorawnative}
\end{figure}

\begin{figure}[p]
\centering
\includegraphics[width={0.45\textwidth}]{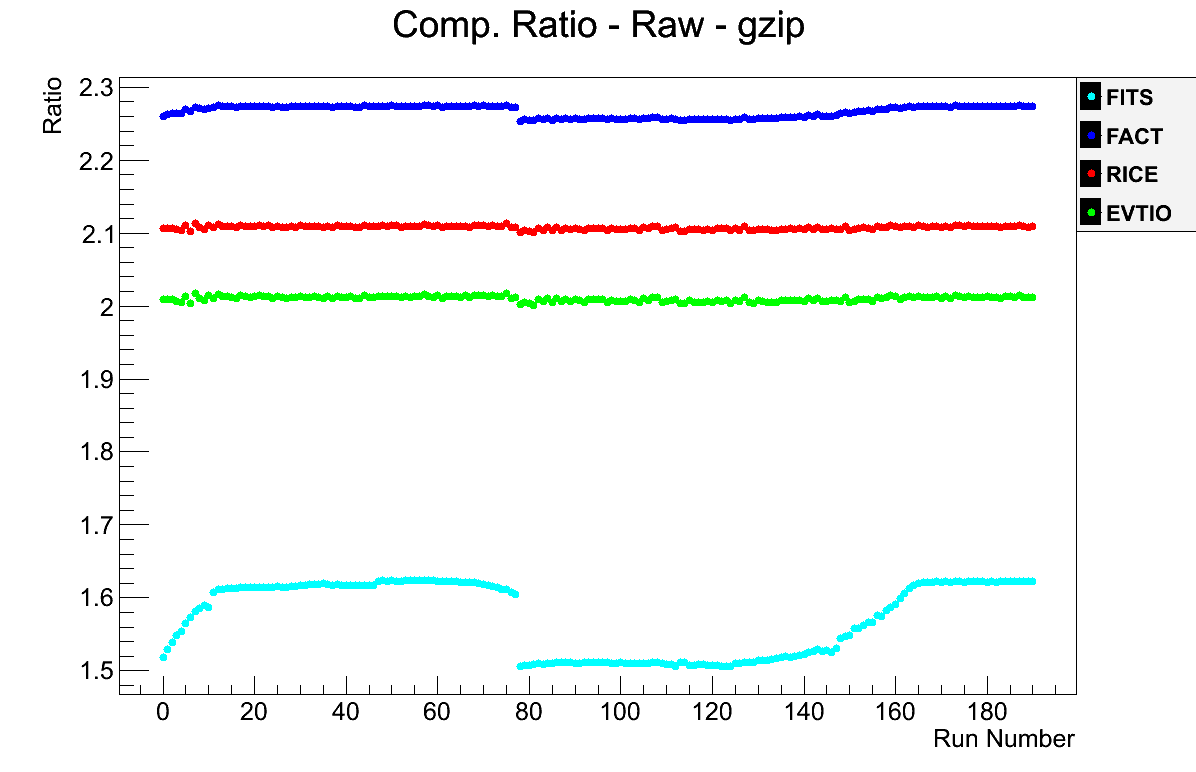}
\includegraphics[width={0.45\textwidth}]{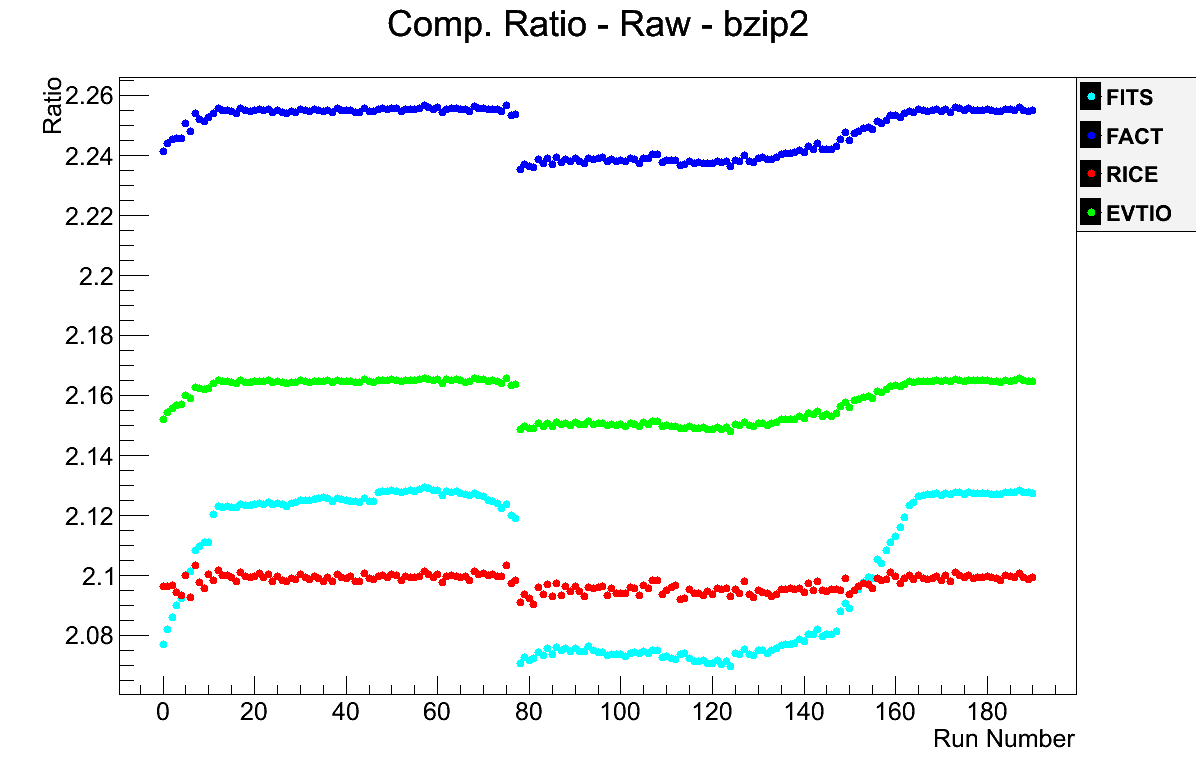}
\caption{Compression ratios of raw data runs for gzip and bzip2 output respectively.}
\label{fig:ratiorawgzip}
\end{figure}

\begin{figure}[p]
\centering
\includegraphics[width={0.45\textwidth}]{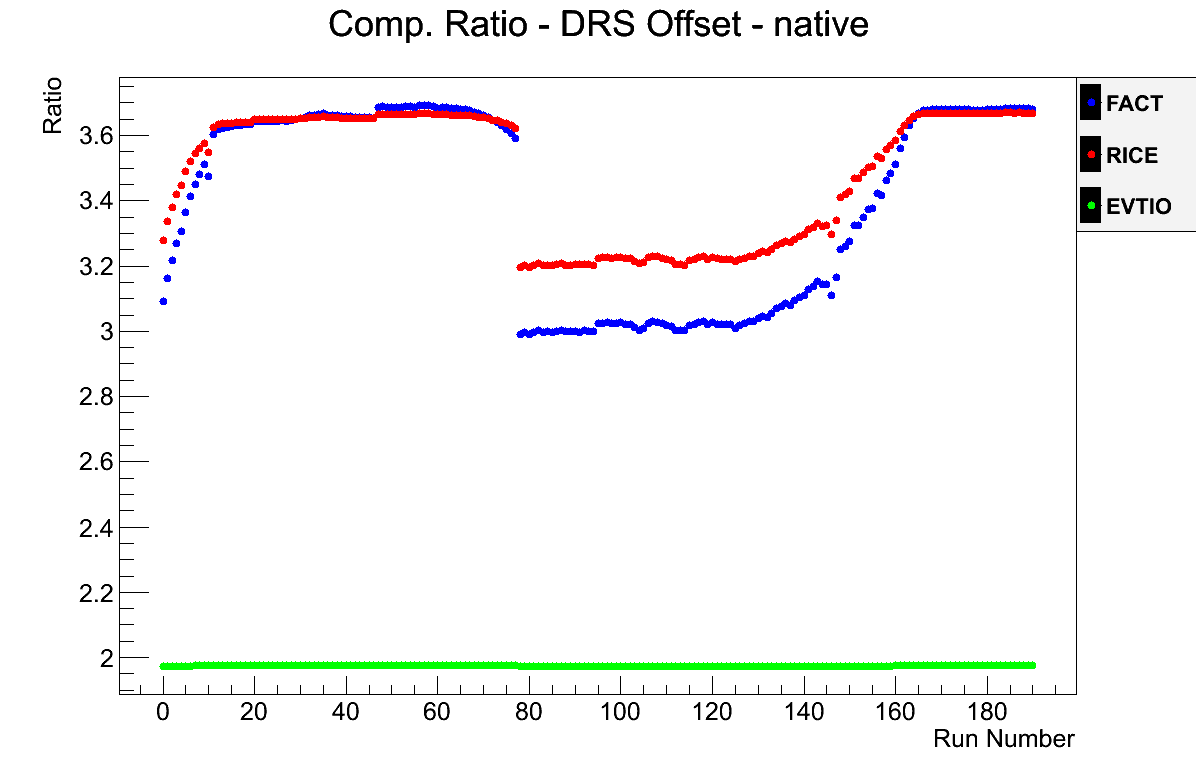}
\includegraphics[width={0.45\textwidth}]{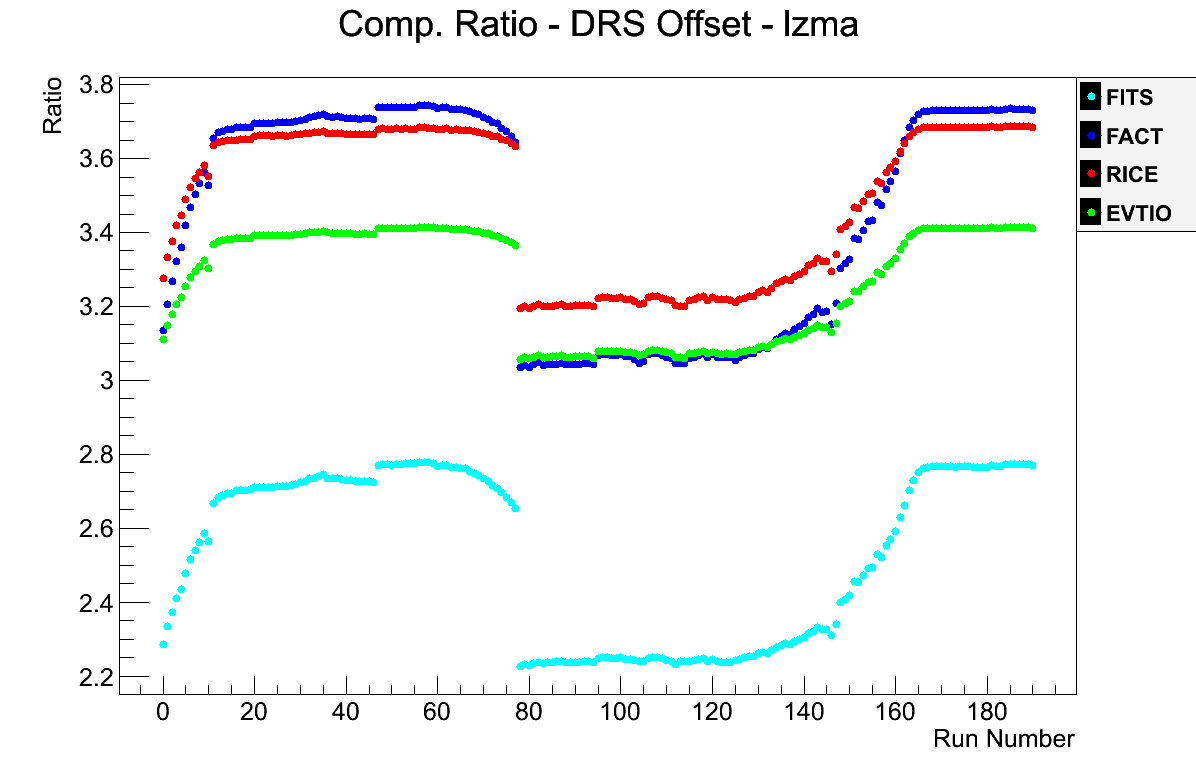}
\caption{Compression ratios of DRS-calibrated data runs for native file format and lzma output respectively.}
\label{fig:ratiocalibnative}
\end{figure}

\begin{figure}[p]
\centering
\includegraphics[width={0.45\textwidth}]{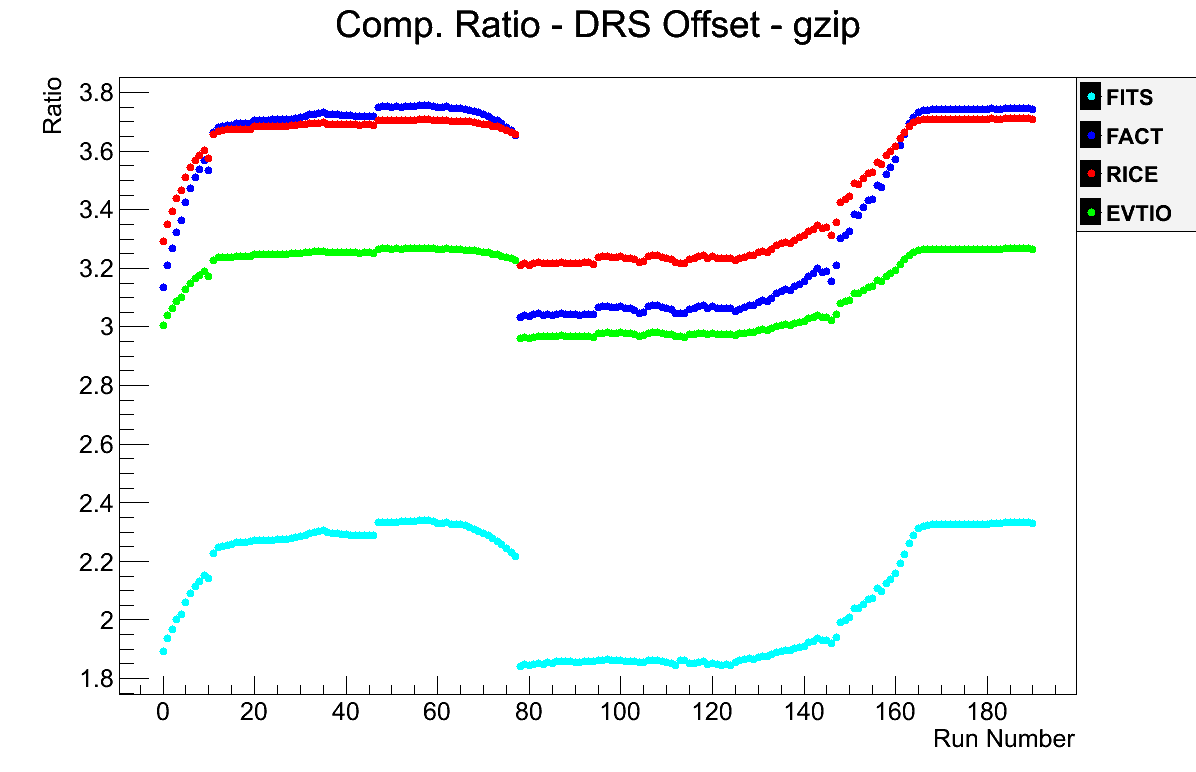}
\includegraphics[width={0.45\textwidth}]{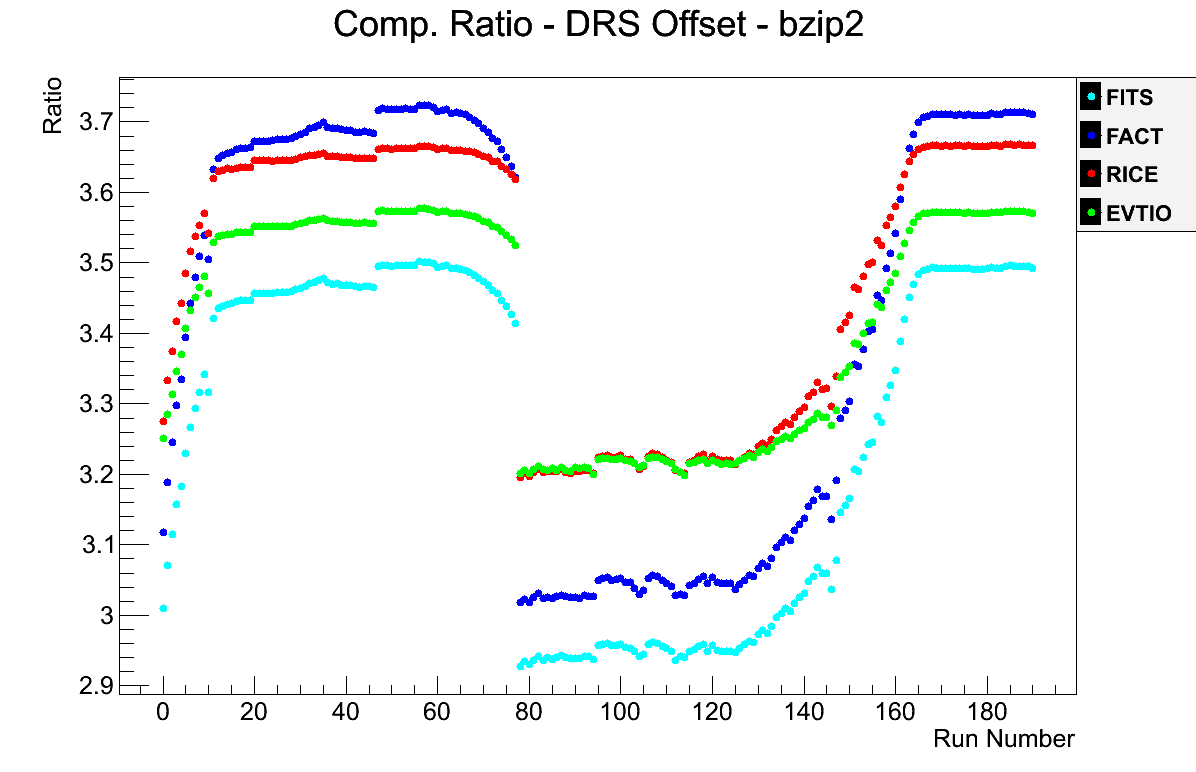}
\caption{Compression ratios of DRS-calibrated data runs for gzip and bzip2 output respectively.}
\label{fig:ratiocalibgzip}
\end{figure}

Figures \ref{fig:ratiorawnative} and \ref{fig:ratiorawgzip} show the compression ratios
obtains for the raw data set. They correspond to the transcoding to the native file format,
lzma, gzip and bzip2 versions respectively. Figures \ref{fig:ratiocalibnative} and 
\ref{fig:ratiocalibgzip} follow the same ordering, only for the DRS-calibrated input data set.

%\begin{figure}[p]
%\centering
%\includegraphics[width={0.95\textwidth}]{plots/comp_ratio_raw_data_lzma.png}
%\caption{Compression ratios of raw data runs for lzma output.}
%\label{fig:ratiorawlzma}
%\end{figure}

%\begin{figure}[p]
%\centering
%\includegraphics[width={0.95\textwidth}]{plots/comp_ratio_raw_data_bzip2.png}
%\caption{Compression ratios of raw data runs for bzip2 output.}
%\label{fig:ratiorawbzip2}
%\end{figure}

%\begin{figure}[p]
%\centering
%\includegraphics[width={0.95\textwidth}]{plots/comp_ratio_calib_data_lzma.png}
%\caption{Compression ratios of drs-offsetted data runs for lzma output.}
%\label{fig:ratiocaliblzma}
%\end{figure}

%\begin{figure}[p]
%\centering
%\includegraphics[width={0.95\textwidth}]{plots/comp_ratio_calib_data_bzip2.png}
%\caption{Compression ratios of drs-offsetted data runs for bzip2 output.}
%\label{fig:ratiocalibbzip2}
%\end{figure}

\subsection{Compression Speed}

\begin{figure}[p]
\centering
\includegraphics[width={0.45\textwidth}]{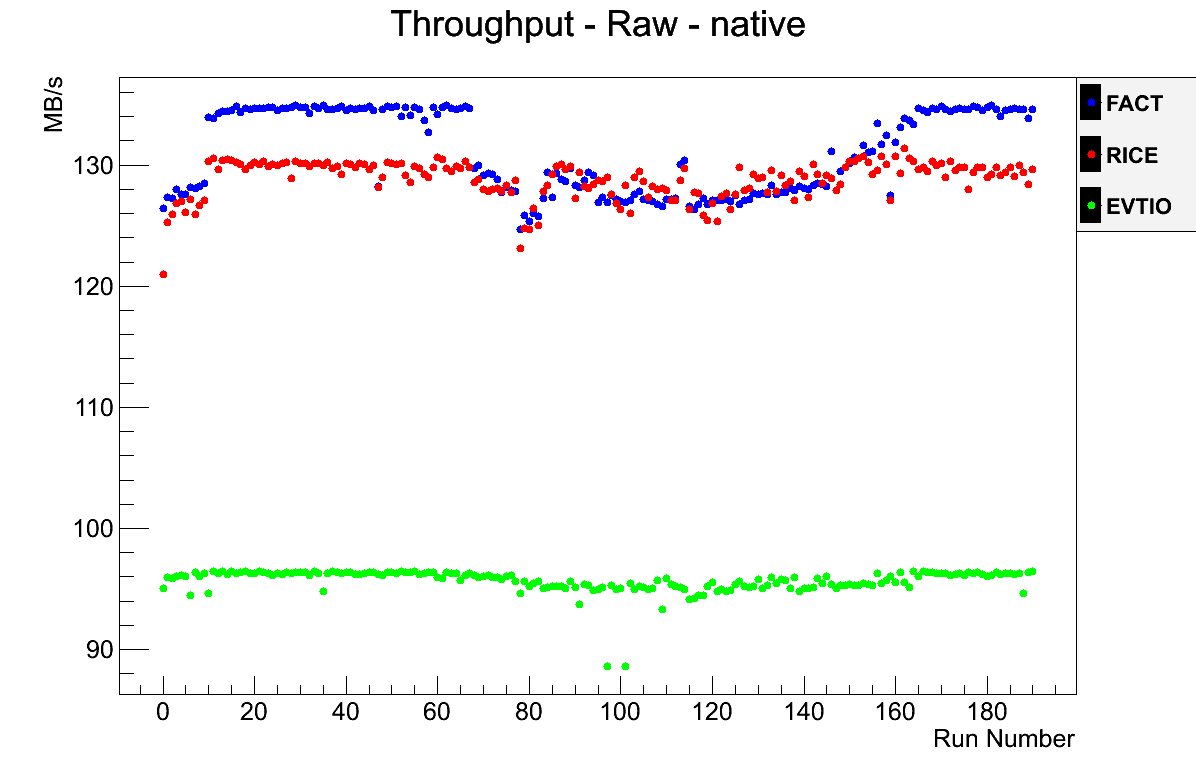}
\includegraphics[width={0.45\textwidth}]{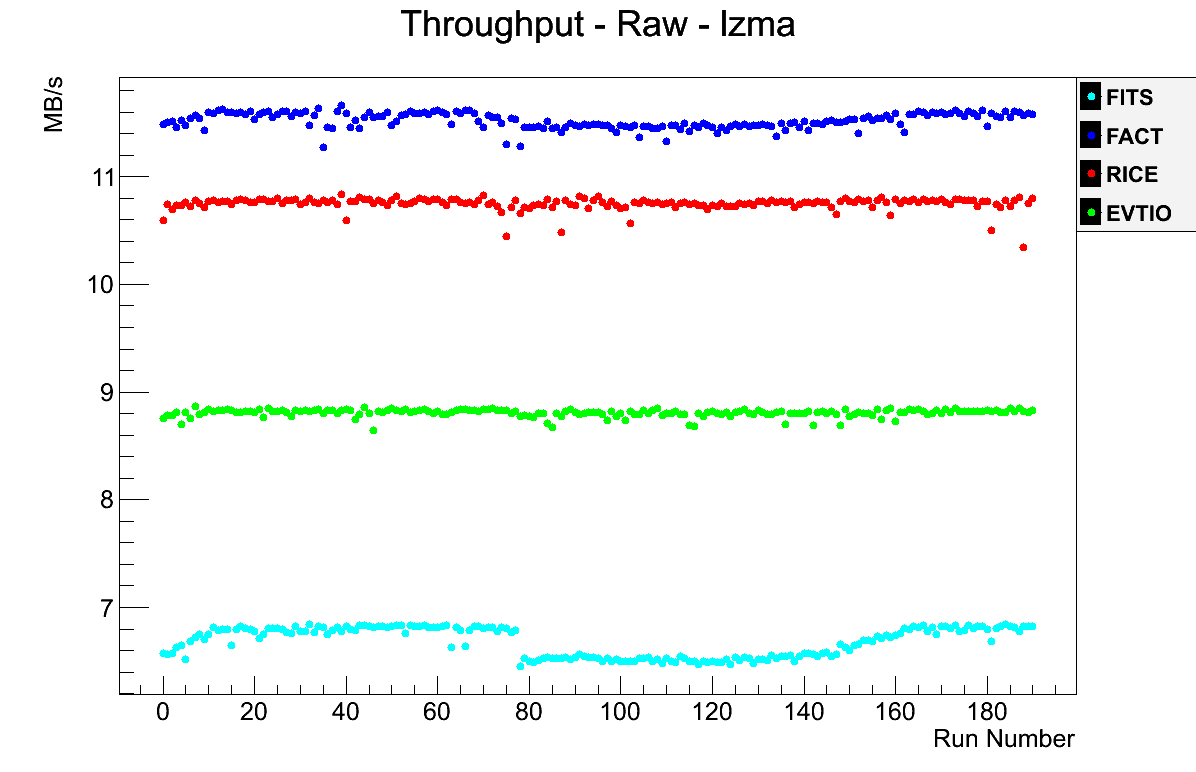}
\caption{Compression throughput in MB/s of raw data runs for native and lzma output respectively.}
\label{fig:mbpsrawnative}
\end{figure}

\begin{figure}[p]
\centering
\includegraphics[width={0.45\textwidth}]{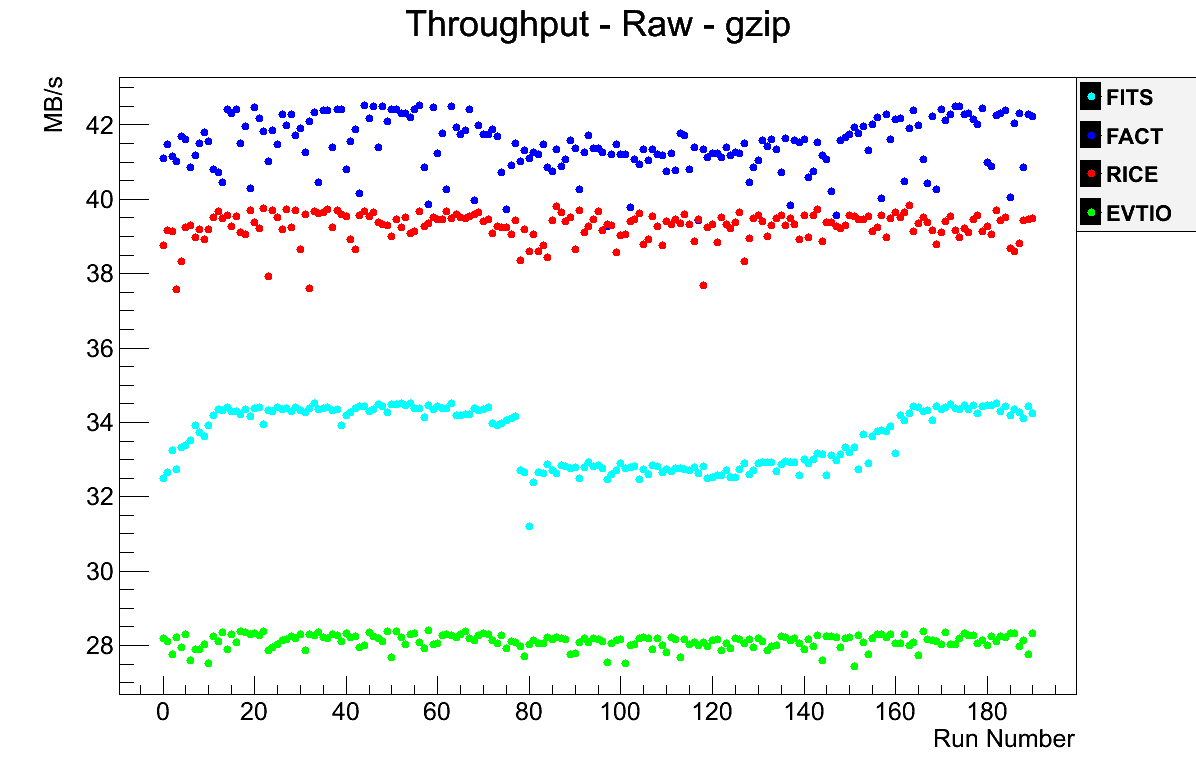}
\includegraphics[width={0.45\textwidth}]{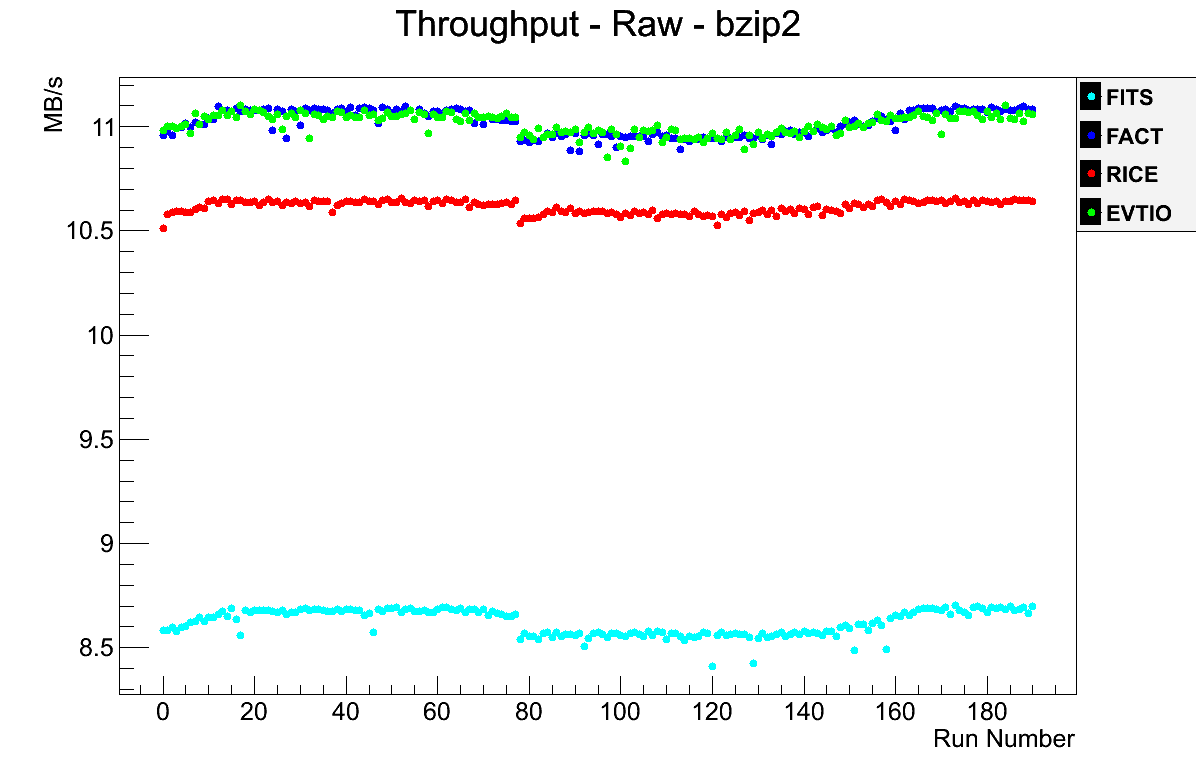}
\caption{Compression throughput in MB/s of raw data runs for gzip and bzip2 output respectively.}
\label{fig:mbpsrawgzip}
\end{figure}

Figures \ref{fig:mbpsrawnative} and \ref{fig:mbpsrawgzip} show the compression speed in MB/s
obtains for the raw data set. They correspond to the transcoding to the native file format,
lzma, gzip and bzip2 versions respectively. 

\begin{figure}[p]
\centering
\includegraphics[width={0.45\textwidth}]{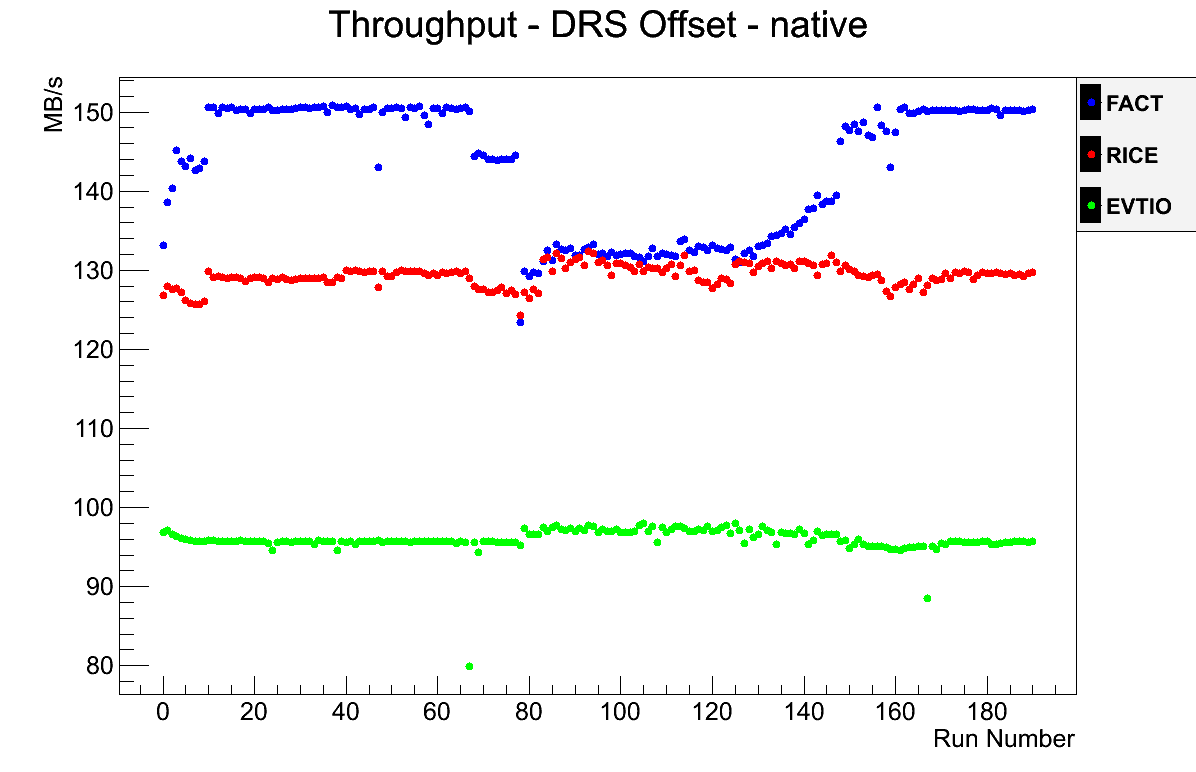}
\includegraphics[width={0.45\textwidth}]{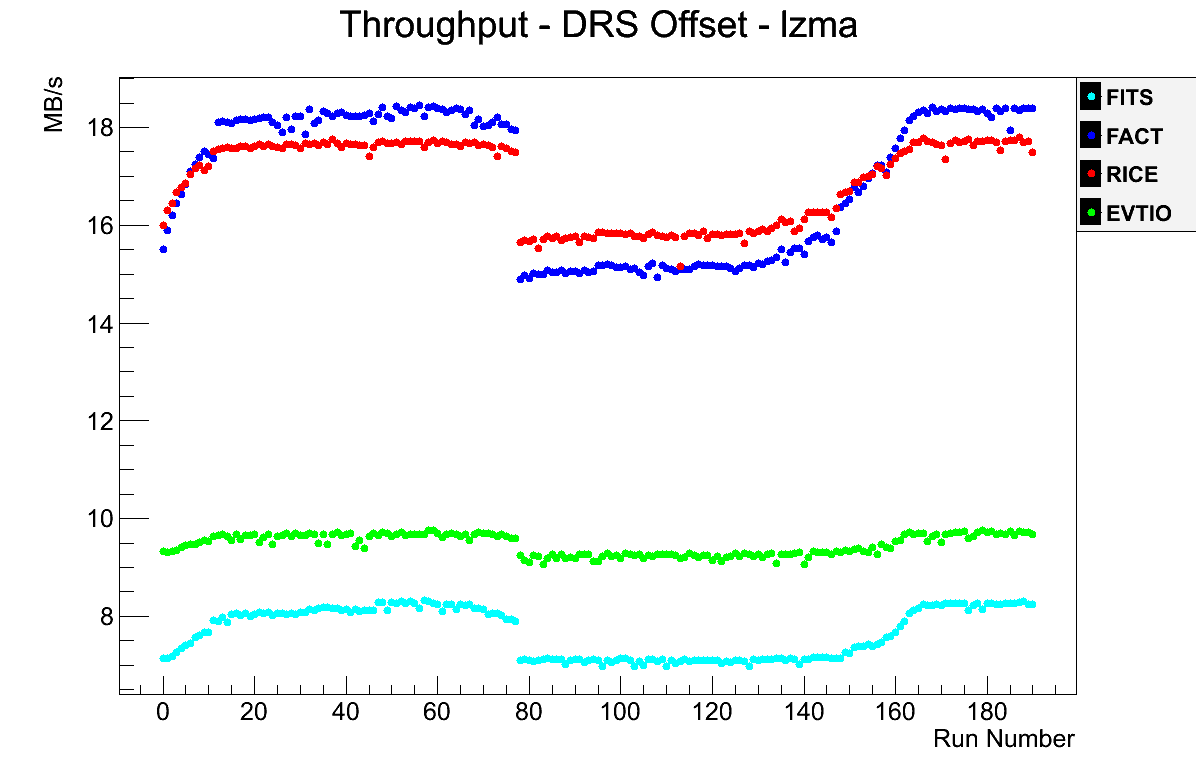}
\caption{Compression throughput in MB/s of DRS-calibrated data runs for native and lzma output respectively.}
\label{fig:mbpscalibnative}
\end{figure}

\begin{figure}[p]
\centering
\includegraphics[width={0.45\textwidth}]{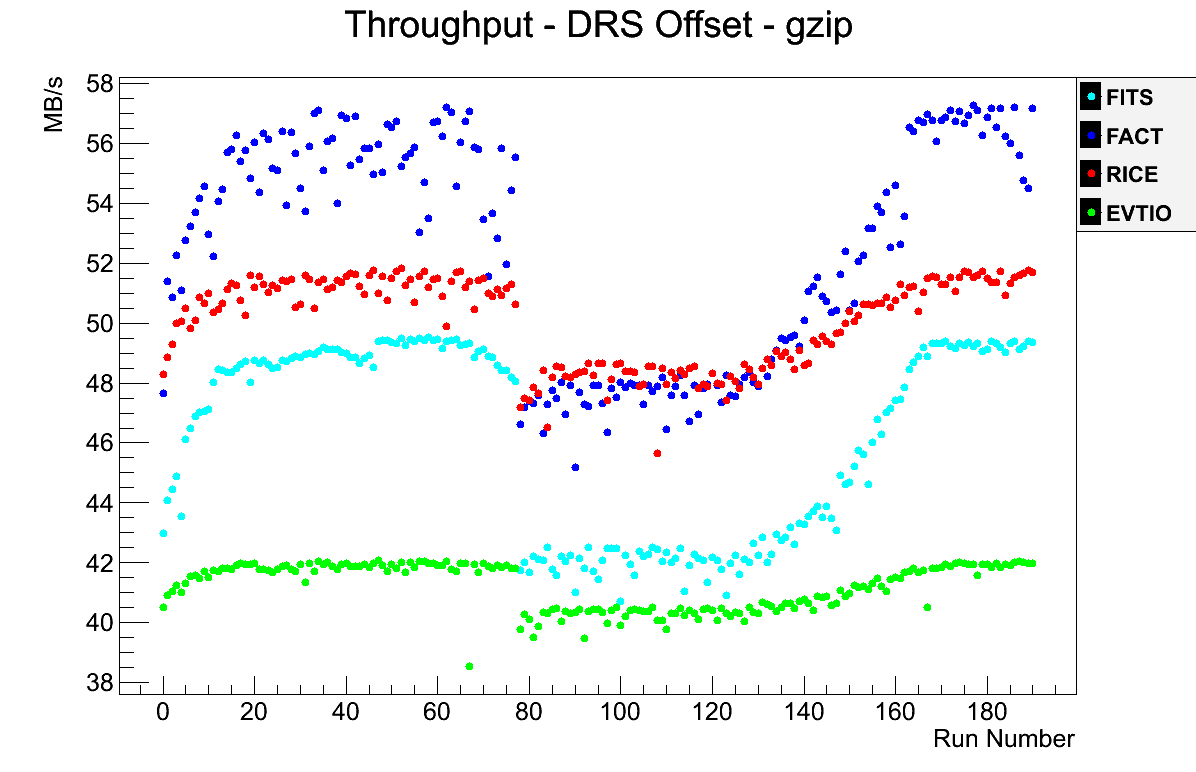}
\includegraphics[width={0.45\textwidth}]{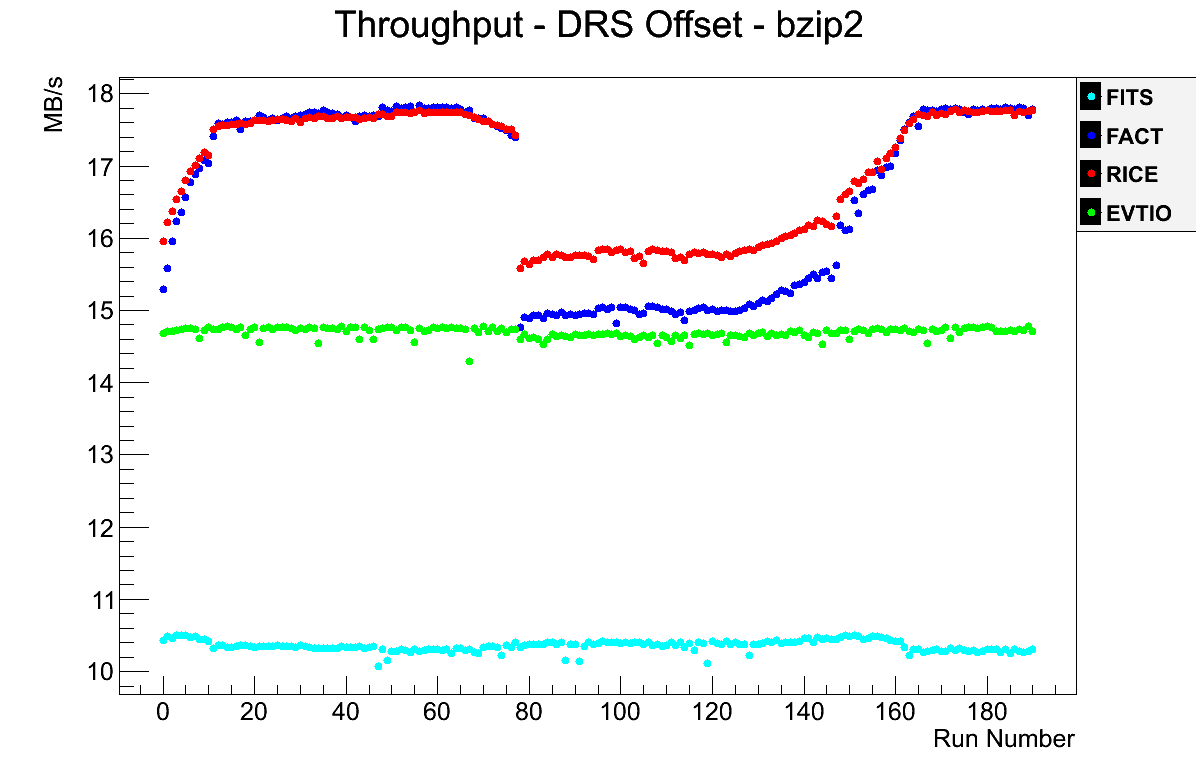}
\caption{Compression throughput in MB/s of DRS-calibrated data runs for gzip and bzip2 output respectively.}
\label{fig:mbpscalibgzip}
\end{figure}

Figures \ref{fig:mbpscalibnative} and \ref{fig:mbpscalibgzip}  
follow the same ordering, only for the DRS-calibrated input data set.

%\begin{figure}[p]
%\centering
%\includegraphics[width={0.95\textwidth}]{plots/comp_speed_raw_data_lzma.png}
%\caption{Compression throughput in MB/s of raw data runs for lzma output.}
%\label{fig:mbpsrawlzma}
%\end{figure}

%\begin{figure}[p]
%\centering
%\includegraphics[width={0.95\textwidth}]{plots/comp_speed_raw_data_bzip2.png}
%\caption{Compression throughput in MB/s of raw data runs for bzip2 output.}
%\label{fig:mbpsrawbzip2}
%\end{figure}

%%%%%%%%%%%%%%%%%%%%%%%%

%\begin{figure}[p]
%\centering
%\includegraphics[width={0.95\textwidth}]{plots/comp_speed_calib_data_lzma.png}
%\caption{Compression throughput in MB/s of drs-offsetted data runs for lzma output.}
%\label{fig:mbpscaliblzma}
%\end{figure}

%\begin{figure}[p]
%\centering
%\includegraphics[width={0.95\textwidth}]{plots/comp_speed_calib_data_bzip2.png}
%\caption{Compression throughput in MB/s of drs-offsetted data runs for bzip2 output.}
%\label{fig:mbpscalibbzip2}
%\end{figure}

\end{appendices}

\end{document}